\documentclass[%
 twocolumn,
 aps,pra,
 amsmath,amssymb,
 superscriptaddress,
 nofootinbib,
 reprint,%
]{revtex4}

\usepackage{stmaryrd}
\usepackage{color}
\usepackage{array}
\usepackage{enumitem}
\usepackage{mathpazo}
\usepackage{setspace}
\usepackage{bm}
\usepackage{amssymb}
\usepackage{amsthm}
\usepackage{mathtools}
\usepackage{booktabs} 

\usepackage{multirow}
\usepackage{cases}

\usepackage[capitalise]{cleveref}
\usepackage{pgfplots}
\usepackage{lipsum}
\usepackage{youngtab}
\usepackage{subfigure}

\newtheorem{definition}{Definition}
\newtheorem{proposition}[definition]{Proposition}
\newtheorem{lemma}[definition]{Lemma}

\newtheorem{theorem}[definition]{Theorem}

\newtheorem{corollary}[definition]{Corollary}
\newtheorem{conjecture}[definition]{Conjecture}

\newtheorem{remark}[definition]{Remark}
\newtheorem{example}[definition]{Example}
\newtheorem{question}[definition]{Question}

\def\Dbar{\leavevmode\lower.6ex\hbox to 0pt
{\hskip-.23ex\accent"16\hss}D}
\makeatletter
\def\url@leostyle{%
  \@ifundefined{selectfont}{\def\UrlFont{\sf}}{\def\UrlFont{\small\ttfamily}}}
\makeatother
\DeclareMathOperator{\tr}{tr} %
\DeclareMathOperator{\Tr}{Tr} %

\def\bcj{\begin{conjecture}}
\def\ecj{\end{conjecture}}
\def\bcr{\begin{corollary}}
\def\ecr{\end{corollary}}
\def\bd{\begin{definition}}
\def\ed{\end{definition}}
\def\bea{\begin{eqnarray}}
\def\eea{\end{eqnarray}}
\def\bem{\begin{enumerate}}
\def\eem{\end{enumerate}}
\def\bex{\begin{example}}
\def\eex{\end{example}}
\def\bim{\begin{itemize}}
\def\eim{\end{itemize}}
\def\bl{\begin{lemma}}
\def\el{\end{lemma}}
\def\bpf{\begin{proof}}
\def\epf{\end{proof}}
\def\bpp{\begin{proposition}}
\def\epp{\end{proposition}}
\def\bqu{\begin{question}}
\def\equ{\end{question}}
\def\br{\begin{remark}}
\def\er{\end{remark}}
\def\bt{\begin{theorem}}
\def\et{\end{theorem}}

\def\btb{\begin{tabular}}
\def\etb{\end{tabular}}

\newcommand{\nc}{\newcommand}

 \nc{\bA}{{\bf A}} \nc{\bB}{{\bf B}} \nc{\bC}{{\bf C}}
 \nc{\bD}{{\bf D}} \nc{\bE}{{\bf E}} \nc{\bF}{{\bf F}}
 \nc{\bG}{{\bf G}} \nc{\bH}{{\bf H}} \nc{\bI}{{\bf I}}
 \nc{\bJ}{{\bf J}} \nc{\bK}{{\bf K}} \nc{\bL}{{\bf L}}
 \nc{\bM}{{\bf M}} \nc{\bN}{{\bf N}} \nc{\bO}{{\bf O}}
 \nc{\bP}{{\bf P}} \nc{\bQ}{{\bf Q}} \nc{\bR}{{\bf R}}
 \nc{\bS}{{\bf S}} \nc{\bT}{{\bf T}} \nc{\bU}{{\bf U}}
 \nc{\bV}{{\bf V}} \nc{\bW}{{\bf W}} \nc{\bX}{{\bf X}}
 \nc{\bZ}{{\bf Z}}

\nc{\cA}{{\cal A}} \nc{\cB}{{\cal B}} \nc{\cC}{{\cal C}}
\nc{\cD}{{\cal D}} \nc{\cE}{{\cal E}} \nc{\cF}{{\cal F}}
\nc{\cG}{{\cal G}} \nc{\cH}{{\cal H}} \nc{\cI}{{\cal I}}
\nc{\cJ}{{\cal J}} \nc{\cK}{{\cal K}} \nc{\cL}{{\cal L}}
\nc{\cM}{{\cal M}} \nc{\cN}{{\cal N}} \nc{\cO}{{\cal O}}
\nc{\cP}{{\cal P}} \nc{\cQ}{{\cal Q}} \nc{\cR}{{\cal R}}
\nc{\cS}{{\cal S}} \nc{\cT}{{\cal T}} \nc{\cU}{{\cal U}}
\nc{\cV}{{\cal V}} \nc{\cW}{{\cal W}} \nc{\cX}{{\cal X}}
\nc{\cZ}{{\cal Z}}

\nc{\hA}{{\hat{A}}} \nc{\hB}{{\hat{B}}} \nc{\hC}{{\hat{C}}}
\nc{\hD}{{\hat{D}}} \nc{\hE}{{\hat{E}}} \nc{\hF}{{\hat{F}}}
\nc{\hG}{{\hat{G}}} \nc{\hH}{{\hat{H}}} \nc{\hI}{{\hat{I}}}
\nc{\hJ}{{\hat{J}}} \nc{\hK}{{\hat{K}}} \nc{\hL}{{\hat{L}}}
\nc{\hM}{{\hat{M}}} \nc{\hN}{{\hat{N}}} \nc{\hO}{{\hat{O}}}
\nc{\hP}{{\hat{P}}} \nc{\hR}{{\hat{R}}} \nc{\hS}{{\hat{S}}}
\nc{\hT}{{\hat{T}}} \nc{\hU}{{\hat{U}}} \nc{\hV}{{\hat{V}}}
\nc{\hW}{{\hat{W}}} \nc{\hX}{{\hat{X}}} \nc{\hZ}{{\hat{Z}}}

\newcommand{\bra}[1]{\langle#1|}
\newcommand{\ket}[1]{|#1\rangle}

\def\Dbar{\leavevmode\lower.6ex\hbox to 0pt
{\hskip-.23ex\accent"16\hss}D}
\begin{document}

%%%%%%%%%%%%%%%%%%%%%%%%%%%%%% added by Muxin %%%%%%%%%%%%%%%%%%%%%%%%%%%%%%%%%%%%%%%%%
%  technical abbreviations

\def\be{\begin{eqnarray}}
\def\ee{\end{eqnarray}}

%%% Calligraphic Alphabet

\newcommand{\ca}{\mathcal A}

\newcommand{\cb}{\mathcal B}
\newcommand{\cc}{\mathcal C}
\newcommand{\cd}{\mathcal D}
\newcommand{\ce}{\mathcal E}
\newcommand{\cf}{\mathcal F}
\newcommand{\cg}{\mathcal G}
\newcommand{\ch}{\mathcal H}
\newcommand{\ci}{\mathcal I}
\newcommand{\cj}{\mathcal J}
\newcommand{\ck}{\mathcal K}
\newcommand{\cl}{\mathcal L}
\newcommand{\cm}{\mathcal M}
\newcommand{\cn}{\mathcal N}
\newcommand{\co}{\mathcal O}
\newcommand{\cp}{\mathcal P}
\newcommand{\cq}{\mathcal Q}
\newcommand{\calr}{\mathcal R}
\newcommand{\cs}{\mathcal S}
\newcommand{\ct}{\mathcal T}
\newcommand{\cu}{\mathcal U}
\newcommand{\cv}{\mathcal V}
\newcommand{\cw}{\mathcal W}
\newcommand{\cx}{\mathcal X}
\newcommand{\cy}{\mathcal Y}
\newcommand{\cz}{\mathcal Z}

%%% script Alphabet

\newcommand{\sa}{\mathscr{A}}
\newcommand{\sm}{\mathscr{M}}

%%% Fraktur Alphabet

\newcommand{\fa}{\mathfrak{a}}  \newcommand{\Fa}{\mathfrak{A}}
\newcommand{\fb}{\mathfrak{b}}  \newcommand{\Fb}{\mathfrak{B}}
\newcommand{\fc}{\mathfrak{c}}  \newcommand{\Fc}{\mathfrak{C}}
\newcommand{\fd}{\mathfrak{d}}  \newcommand{\Fd}{\mathfrak{D}}
\newcommand{\fe}{\mathfrak{e}}  \newcommand{\Fe}{\mathfrak{E}}
\newcommand{\ff}{\mathfrak{f}}  \newcommand{\Ff}{\mathfrak{F}}
\newcommand{\fg}{\mathfrak{g}}  \newcommand{\Fg}{\mathfrak{G}}
\newcommand{\fh}{\mathfrak{h}}  \newcommand{\Fh}{\mathfrak{H}}
\newcommand{\fraki}{\mathfrak{i}}       \newcommand{\Fraki}{\mathfrak{I}}
\newcommand{\fj}{\mathfrak{j}}  \newcommand{\Fj}{\mathfrak{J}}
\newcommand{\fk}{\mathfrak{k}}  \newcommand{\Fk}{\mathfrak{K}}
\newcommand{\fl}{\mathfrak{l}}  \newcommand{\Fl}{\mathfrak{L}}
\newcommand{\fm}{\mathfrak{m}}  \newcommand{\Fm}{\mathfrak{M}}
\newcommand{\fn}{\mathfrak{n}}  \newcommand{\Fn}{\mathfrak{N}}
\newcommand{\fo}{\mathfrak{o}}  \newcommand{\Fo}{\mathfrak{O}}
\newcommand{\fp}{\mathfrak{p}}  \newcommand{\Fp}{\mathfrak{P}}
\newcommand{\fq}{\mathfrak{q}}  \newcommand{\Fq}{\mathfrak{Q}}
\newcommand{\fr}{\mathfrak{r}}  \newcommand{\Fr}{\mathfrak{R}}
\newcommand{\fs}{\mathfrak{s}}  \newcommand{\Fs}{\mathfrak{S}}
\newcommand{\ft}{\mathfrak{t}}  \newcommand{\Ft}{\mathfrak{T}}
\newcommand{\fu}{\mathfrak{u}}  \newcommand{\Fu}{\mathfrak{U}}
\newcommand{\fv}{\mathfrak{v}}  \newcommand{\Fv}{\mathfrak{V}}
\newcommand{\fw}{\mathfrak{w}}  \newcommand{\Fw}{\mathfrak{W}}
\newcommand{\fx}{\mathfrak{x}}  \newcommand{\Fx}{\mathfrak{X}}
\newcommand{\fy}{\mathfrak{y}}  \newcommand{\Fy}{\mathfrak{Y}}
\newcommand{\fz}{\mathfrak{z}}  \newcommand{\Fz}{\mathfrak{Z}}

\newcommand{\cfg}{\dot \fg}
\newcommand{\cFg}{\dot \Fg}
\newcommand{\ccg}{\dot \cg}
\newcommand{\circj}{\dot {\mathbf J}}
\newcommand{\circs}{\circledS}
\newcommand{\jmot}{\mathbf J^{-1}}

%%% Greek letters

\newcommand{\rmd}{\mathrm d}
\newcommand{\mca}{\ ^-\!\!\ca}
\newcommand{\pca}{\ ^+\!\!\ca}
\newcommand{\peq}{^\Psi\!\!\!\!\!=}
\newcommand{\lt}{\left}
\newcommand{\rt}{\right}
\newcommand{\HN}{\hat{H}(N)}
\newcommand{\HM}{\hat{H}(M)}
\newcommand{\Hv}{\hat{H}_v}
\newcommand{\cyl}{\mathbf{Cyl}}
\newcommand{\lag}{\left\langle}
\newcommand{\rag}{\right\rangle}
\newcommand{\Ad}{\mathrm{Ad}}
\newcommand{\trace}{\mathrm{tr}}
\newcommand{\bbc}{\mathbb{C}}
\newcommand{\AC}{\overline{\mathcal{A}}^{\mathbb{C}}}
\newcommand{\Ar}{\mathbf{Ar}}
\newcommand{\uc}{\mathrm{U(1)}^3}
\newcommand{\M}{\hat{\mathbf{M}}}
\newcommand{\spin}{\text{Spin(4)}}
\newcommand{\id}{\mathrm{id}}
\newcommand{\Pol}{\mathrm{Pol}}
\newcommand{\Fun}{\mathrm{Fun}}
\newcommand{\bp}{p}
\newcommand{\act}{\rhd}
\newcommand{\data}{\lt(j_{ab},A,\bar{A},\xi_{ab},z_{ab}\rt)}
\newcommand{\datao}{\lt(j^{(0)}_{ab},A^{(0)},\bar{A}^{(0)},\xi_{ab}^{(0)},z_{ab}^{(0)}\rt)}
\newcommand{\deltadata}{\lt(j'_{ab}, A',\bar{A}',\xi_{ab}',z_{ab}'\rt)}
\newcommand{\background}{\lt(j_{ab}^{(0)},g_a^{(0)},\xi_{ab}^{(0)},z_{ab}^{(0)}\rt)}
\newcommand{\sgn}{\mathrm{sgn}}
\newcommand{\vth}{\vartheta}
\newcommand{\rmi}{\mathrm{i}}
\newcommand{\bfmu}{\pmb{\mu}}
\newcommand{\bfnu}{\pmb{\nu}}
\newcommand{\bfm}{\mathbf{m}}
\newcommand{\bfn}{\mathbf{n}}
\newcommand{\perk}{\mathfrak{S}_k}
\newcommand{\dens}{\mathrm{D}}
\newcommand{\iden}{\mathbb{I}}
\newcommand{\End}{\mathrm{End}}
\newcommand{\C}{\mathbb{C}}

%%%%%%%%%%%%%%% mathscr

\newcommand{\sz}{\mathscr{Z}}
\newcommand{\sk}{\mathscr{K}}

\title{Upper bounds for relative entropy of entanglement based on active learning}

\author{Shi-Yao Hou}%
\affiliation{College of Physics and Electronic Engineering, Center for Computational Sciences,  Sichuan Normal University, Chengdu 610068, China}
\affiliation{Department of Physics, The Hong Kong University of Science and Technology, Clear Water Bay, Kowloon, Hong Kong, China}
%\affiliation{Center for Quantum Computing, Peng Cheng Laboratory, Shenzhen, 518055, China}
\author{Chenfeng Cao}
\affiliation{Department of Physics, The Hong Kong University of Science and Technology, Clear Water Bay, Kowloon, Hong Kong, China}

\author{D. L. Zhou}
\affiliation{Institute
  of Physics, Beijing National Laboratory for Condensed Matter
  Physics, Chinese Academy of Sciences, Beijing 100190, China}
\affiliation{School of Physical Sciences, University of Chinese
  Academy of Sciences, Beijing 100049, China}
\affiliation{CAS Central of Excellence in Topological Quantum
  Computation, Beijing 100190, China}
\affiliation{Songshan Lake Materials Laboratory, Dongguan, Guangdong
  523808, China}

\author{Bei Zeng}
\email[]{zengb@ust.hk}
\affiliation{Department of Physics, The Hong Kong University of Science and Technology, Clear Water Bay, Kowloon, Hong Kong, China}
%\affiliation{Center for Quantum Computing, Peng Cheng Laboratory, Shenzhen, 518055, China}
%\affiliation{Department of Mathematics \& Statistics, University of Guelph, Guelph, Ontario, Canada}%
%\affiliation{Institute for Quantum Computing, University of Waterloo, Waterloo, Ontario, Canada}

\date{\today}% It is always \today, today,
             %  but any date may be explicitly specified

\begin{abstract}
  Quantifying entanglement for multipartite quantum state is a crucial
  task in many aspects of quantum information theory. Among all
  the entanglement measures, relative entropy of entanglement $E_{R}$ is an
  outstanding quantity due to its clear geometric meaning, easy
  compatibility with different system sizes, and various applications in 
  many other related quantity calculations. Lower bounds of $E_R$ were
  previously found based on distance to the set of positive partial
  transpose states. We propose a method to calculate upper bounds of
  $E_R$ based on active learning, a subfield in machine learning,
  to generate an approximation of the set of separable
  states. We apply our method to calculate $E_R$ for composite systems
  of various sizes, and compare with the previous known lower bounds,
  obtaining promising results. Our method adds a reliable tool for
  entanglement measure calculation and deepens our understanding for
  the structure of separable states.
\end{abstract}

\maketitle
\renewcommand\theequation{\arabic{section}.\arabic{equation}}
\setcounter{tocdepth}{4}
%\tableofcontents
\makeatletter
\@addtoreset{equation}{section}
\makeatother

\section{Introduction}

Quantum entanglement is a kind of correlation that is beyond any
possible classical probabilistic correlation
~\cite{horodecki2009quantum}. Entanglement has played a crucial role
in almost all aspects of quantum information theory, such as quantum
channel capacity~\cite{lloyd1997capacity}, quantum
algorithm~\cite{farhi2001quantum}, quantum error
correction~\cite{bennett1996mixed} and quantum
sensing~\cite{huver2008entangled}. In recent years, entanglement
also becomes a central concept in the interdisciplinary field of
quantum information theory, condensed matter physics, and quantum
gravity~\cite{zeng2019quantum,almheiri2015bulk}.

For any given (multipartite) quantum state $\rho$, one fundamental
question one would like to know is whether $\rho$ is entangled or not,
and a further question, how much $\rho$ is
entangled~\cite{guhne2009entanglement}. Entanglement measures are
quantities providing such kind of information. Normally, an
entanglement measure $E(\rho)$ satisfies some natural assumptions such as
invariance under local unitary operations, and non-increasing under general
local operations~\cite{vedral1997quantifying,vedral1998entanglement}.

\begin{figure}
\includegraphics[scale=0.5]{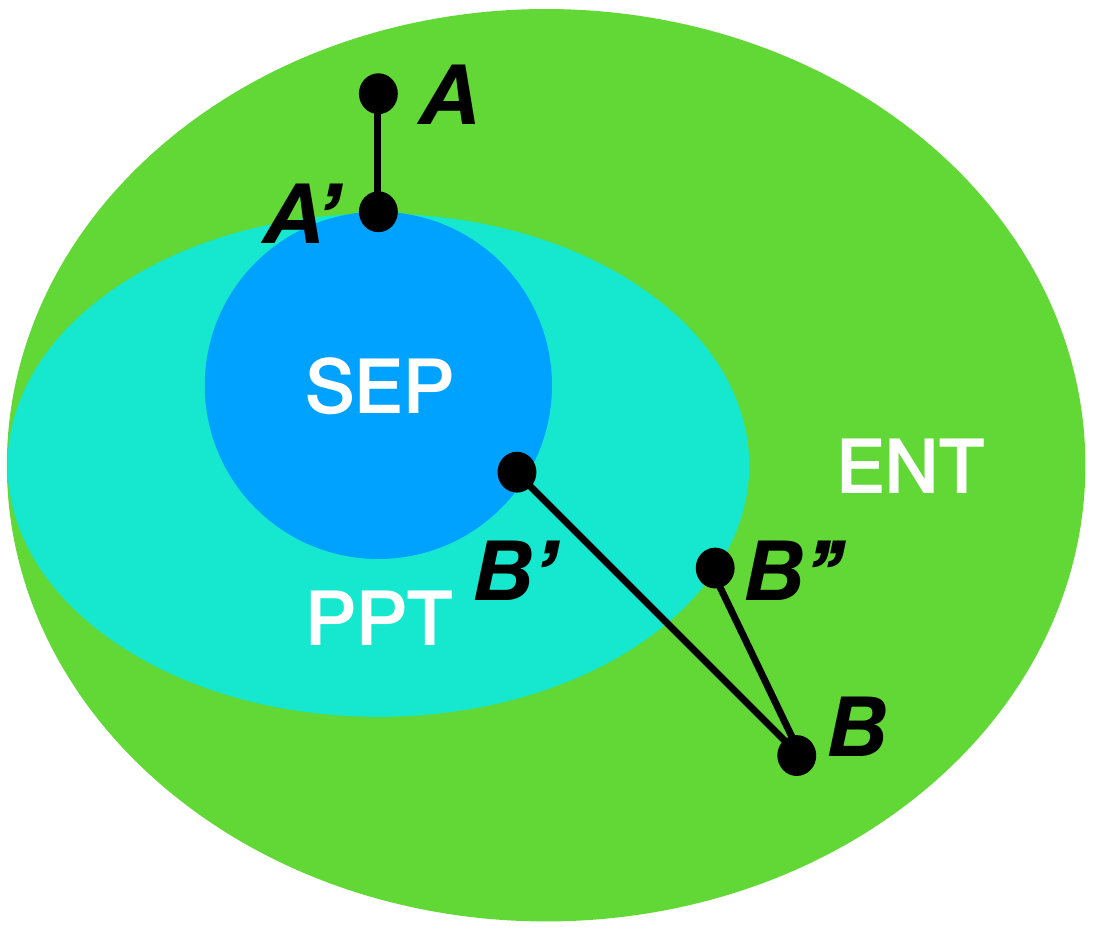}
\caption{Demonstration of the quantum state space. SEP represents the set of separable states, PPT represents the set of states with positive partial transpose, and ENT represents the set of entangled states. 
%Since the quantum state space is a high dimensional space, this is only a demonstration. 
$A$ and $B$ represents the state $\rho$ for which the relative entropy of entanglement is to be calculated. For the state $A$, the boundary state $\sigma$ satisfies the PPT criterion. However, for the state $B$, the boundary state $B'$ does not satisfy the PPT criterion. Therefore, the relative entropy of entanglement calculated using PPT states is smaller than the actual relative entropy of entanglement.}
\label{fig:demo}
\end{figure}

Among many entanglement measures, relative entropy of entanglement
$E_R$ is one important quantity~\cite{vedral2002role}. For any quantum
state $\rho$, $E_R(\rho)$ naturally measures ``how far" $\rho$ is from the set
of separable states. As illustrated in \cref{fig:demo}, inside the quantum
state space, for an entangled state, which is represented by $A$ or
$B$, the relative entropy of entanglement is the \textit{distance} of
the state to the set of separable states (denoted by SEP in the
figure). Or equivalently, the task of finding $E_R(\rho)$ is to look for 
a state in SEP ($A'$ or $B'$) to
minimize \textit{distance} to the entangled state $\rho$ ($A$ or $B$).
Besides having a good geometric interpretation, $E_R(\rho)$ is also
known to be compatible for multiparty systems, providing an upper
bound for entanglement distillation
~\cite{horodecki2000limits,rains2001semidefinite}, and connected to
the study of many other aspects in quantum information theory, such as
the use of relative entropy in some information-theoretic
quantities~\cite{vedral2002role,henderson2000information}.

Calculation of $E_R(\rho)$ is an optimization problem over the set of
separable states~\cite{girard2014convex}. If you can characterize the set, then the calculation of $E_R(\rho)$ can be solved by semi-definite programming. However, such a set of separable
states is notoriously hard to characterize, despite known to be
convex~\cite{horodecki2009quantum}. On the other hand, the set of
states with positive partial transpose (PPT) is much easier to
characterize~\cite{horodecki1997separability}. Calculating ``how far''
the state $\rho$ is from the set of PPT states in terms of relative entropy
can then be formulated as a certain kind of Semi-Definite Programming
(SDP)~\cite{Fawzi_2018,miranowicz2004comparative}. This then gives a
lower bound of $E_R(\rho)$, as illustrated in \cref{fig:demo}. As we
know, the set of separable states (SEP in \cref{fig:demo}) is a subset
of the set of PPT states (PPT in \cref{fig:demo}). Therefore, for some states,
such as the state $B$, the closest state in PPT is $B"$, while the
closest state in SEP is $B'$. Consequently, the \textit{distance} of $BB'$ is
smaller than $BB"$, hence the point $B"$ gives a lower bound for $E_R(\rho)$.

%Active learning, also called query learning, or optimal experiment design, is one subfield in machine learning. 
%Most supervised learning algorithms require a large amount of data. For any supervised learning algorithm to work well, a large amount of labeled data is required. However, for many problems, the data are unlabelled, or labelling of the data is an expensive task. As a prior knowledge, we know some points on the boundary of the set of separable states.  The linear combination of some of these points gives an approximation of the relative entropy of entanglement. How accurate the approximation is depends on the sampling of those points. As we will see later, some of those points on the boundary are \textit{useful} and some are \textit{useless}. Whether a point is useful or useless depends on the actual given state $\rho$, hence unlabelled. We will use active learning to select these points and approximate the set of all separable states.

In order to get an upper bound of $E_R(\rho)$, we will need to approximate the set of separable states from inside. 
Notice that the extreme points of the set
of separable states, which is convex, are simply pure product states. 
Intuitively, we can use, e.g. the convex hull approximation (CHA) as described in~\cite{sirui2018ent} to generate 
such an approximation. However, since the set of separable states has a much larger dimension than
the set of its extreme points, how to select the extreme points is a difficult task -- simple random sampling 
does not lead to good approximation~\cite{sirui2018ent}. 

To overcome the above-mentioned difficulty, we propose to use the a method based on active learning. Active learning, also called query learning, or optimal experiment design, is one subfield in machine learning.  In general, most supervised learning algorithms require a large amount of data. However, for many problems, the data are unlabelled, or labelling of the data is an expensive task. Actively learning can be used to dynamically label useful data during the training process~\cite{castro2008minimax, freund1997selective, balcan2009agnostic}. In our case, as a prior knowledge, we know some points on the boundary of the set of separable states e.g. we can start from some extreme points and some linear combination of them.  How accurate the approximation depends on the sampling of those points. In other words, some of those points on the boundary are \textit{useful} and some are \textit{useless}. Whether a point is useful or useless depends on the actual state $\rho$ given, hence unlabelled. We can then use active learning to select these points and improve our approximation to the set of all separable states.

We apply our method to calculate $E_R(\rho)$ for composite systems of
various sizes, and compare our results with the previous lower bounds given by
SDP. We consider bipartite system of dimension $d_A\otimes d_B$. In the case of
$d_Ad_B\leq 6$, where the set of PPT is the same as that of SEP~\cite{horodecki1997separability}, 
our results are very close to the former results based on PPT. In the case of
$d_Ad_B>6$, our method gives an upper bound, which is always 
larger that the value given by PPT, and in many cases we believe that our value is closer to the actual value of $E_R(\rho)$. Our results add
a new tool for entanglement measure calculation and deepen our
understanding about the difference between the set structures of
SEP and PPT.

We organize our paper as follows: in Sec II, we discuss our method based on active
learning. In Sec III, we show our results in different situations: 
firstly for two kinds of states (Werner states and isotropic states),
for which we know the analytical form of their relative entropy of
entanglement, to compare and check the validity of our algorithm; 
secondly for a special case of states that are bound entangled 
(i.e. entangled state that
are PPT),
we calculate their relative entropy of
entanglement, to demonstrate that our method
gives better estimation for the true value of the method of PPT; finally for different dimensions, we
generate random states and calculate their relative entropy of
entanglement, to demonstrate the power of our method
for giving new understandings of the difference between SEP and PPT. In Sec IV, 
we summarize our results and discuss some future directions.

\section{Method based on active learning}

In this section, we discuss our algorithm for calculating the upper
bounds of relative entropy of entanglement based on active learning.
We start to recall some basic properties of $E_R(\rho)$, then discuss 
the active learning method and its application for calculating
the upper bound of $E_R(\rho)$.

\subsection{Relative entropy of entanglement}

For a given bipartite system $AB$ with Hilbert
space $\mathcal{H}_A\otimes\mathcal{H}_B$, a state $\rho_{AB}$ is separable if
it can be written in a convex combination
\begin{equation}
\rho_{AB}=\sum_i \lambda_i \rho_A^{(i)} \otimes \rho_B^{(i)},
\label{eq:ent}
\end{equation}
where $\rho_A$ and $\rho_B$ are states in Hilbert space $\mathcal{H}_A$ and
$\mathcal{H}_B$, respectively. The set of all separable
states, denoted by SEP, is a convex set, given by its definition. 
The extreme points of SEP are given by the states of the form
$\rho_A\otimes\rho_B$, where $\rho_A$ and $\rho_B$ are arbitrary pure states. 

Denote the convex set of SEP as
$\mathcal{D}$. For any given bipartite state $\rho$, 
the relative entropy of entanglement $E_R(\rho)$ is given by finding
a state $\tau\in\mathcal{D}$ that minimizes  
the relative entropy between $\rho$ and $\tau$. That is,
\begin{equation}
E_R(\rho)=\min_{\sigma} S(\rho || \sigma)=\min_{\sigma\in \mathcal D}\Tr(\rho \log \rho -\rho \log \sigma)
\end{equation}
where $\sigma$ is a state in $\mathcal{D}$. However, finding $\sigma$ to
minimize $S(\rho || \sigma)$ is a difficult task, mainly due to the fact that
characterizing the set of SEP is hard~\cite{gurvits2003classical,slater2019}.
In practice, even for systems with
dimension $d_A d_B>6$, it is already hard to find 
whether a given state is separable or not~\cite{johnston2014detection}.

$E_R(\rho)$ could be seen as a
\textit{distance} between the state $\rho$ and the set of SEP.
Finding the \textit{distance} means finding a point from the set which
is the closest to the state. As shown in \cref{fig:demo}, for a
state $\rho_{A}$ represented by the point $A$, the task is to
find the point $A'$ (representing the state $\sigma_{A'}$), to minimize
\begin{equation}
E_R(\rho_{A})=\Tr(\rho\log\rho_{A}-\rho\log\sigma_{A'}).
\end{equation}
With respect to the set of separble states $\mathcal{D}$, the problem could be formulated as the following problem.
\begin{align}
\min  \Tr(\rho(\log\rho-\log\sigma)) \quad \text{s.t} \quad \sigma \in\mathcal{D}, \\
\text{i.e.}\quad\sigma=\sum_i \lambda_i \sigma_i,\sum_i \lambda_i = 1,\lambda_i\geq 0,
\end{align}
where $\sigma_i$s are product states.
Notice that the function \textsf{quantum\_rel\_entr(X,Y)} provided in CVXQUAD~\citep{cvxquad} can give a good approximation of the
true minimum of the relative entropy. Therefore, the above-mentioned optimization problem could be solved using semi-definite programming (SDP). We can use, e.g. the CVX package to perform the task of SDP~\cite{cvx,gb08}.

However, the characterization of the set of SEP is known to be very
difficult, so does the calculation of $E_R(\rho)$.
It has been proposed to approximates the set of SEP
by the set of states with positive partial transpose (denoted by PPT) ~\citep{Fawzi_2018},
which is easier to characterize.
However, it is well known that the set PPT is strictly larger than the set of SEP for
$d_A d_B > 6$. Hence the method based on PPT  gives a lower bound for $E_R(\rho)$,
and does not give any information for bound entangled states.
In order to obtain a better approximation for $E_R(\rho)$, one would then need
a better approximation to the set of SEP from the inside. 

\subsection{Active learning}

Supervised learning has achieved great success in the past decades. However, sometimes only a small part or even none in the data set is labelled due to high labelling cost. Active learning method can dynamically label useful data during the training process. Take classification task as an example, suppose we have a small labeled dataset and a large unlabeled dataset. The algorithm is as follows:

\begin{itemize}
\item[1] Train the classifier with the labelled dataset. If the full dataset is unlabelled, using the oracle to label this dataset first and train the classifier.
\item[2] Calculate the final accuracy.
\item[3] Select data points close to the boundary from the unlabelled dataset, label these samples through an oracle and add it to the labelled training dataset.
\item[4] Repeat step 1 to 3 until it reaches the desired accuracy.
\end{itemize}

In some scenarios, active learning requires only $O(log(k))$ labeled data samples to achieve the same performance as passive learning with $k$ labeled samples \cite{castro2008minimax, freund1997selective, balcan2009agnostic}. We can therefore expect an exponential speed-up of active learning compared with passive learning considering the similarity between our problem and those cases. As the quantum system size $n$ increase, we need to sample exponential random product states to estimate the upper bounds of relative entropy of entanglement. But for active learning, $O(n)$ samples is probably enough. Therefore,  the method based on active learning is scalable and can be applied to larger systems.

\subsection{Algorithm based on active learning}

We now develop an algorithm to estimate the set of separable states. 
We know that the extreme points of SEP are given by 
all the pure product states of the form
\begin{equation}
\rho_{sp}=|i\rangle\langle i|\otimes |j\rangle\langle j|
\end{equation}
with $|i\rangle$ and
$|j\rangle$ are arbitrary pure states in $\mathcal{H}_A$ and $\mathcal{H}_B$,
respectively. We then randomly sample a set of these extreme points
$\{\sigma_{1},\sigma_{2},...,\sigma_{n}\}$, then form a convex hull
$\mathcal{C}$. $\mathcal{C}$ then gives an approximation (i.e. the convex hull approximation, CHA) of $\mathcal{D}$ from 
inside. Thus the result is an upper bound. The result will be in the form of 
\begin{equation}
\sigma=\sum_i c_i \sigma_i.
\end{equation}

Intuitively, we can then build a classifier based on CHA using supervised learning by constructing and training a deep neural network (DNN), to characterize the set of separable states. 
However, this does not result in a good approximation as discussed in~\cite{sirui2018ent}. Or in other words,
simple random sampling does not lead to good approximation. 
This is due to the fact that the set of separable states has a much larger dimension than
the set of its extreme points. Consequently, how to select the extreme points becomes a difficult task. 

\begin{figure}[!ht]
\includegraphics[scale=0.5]{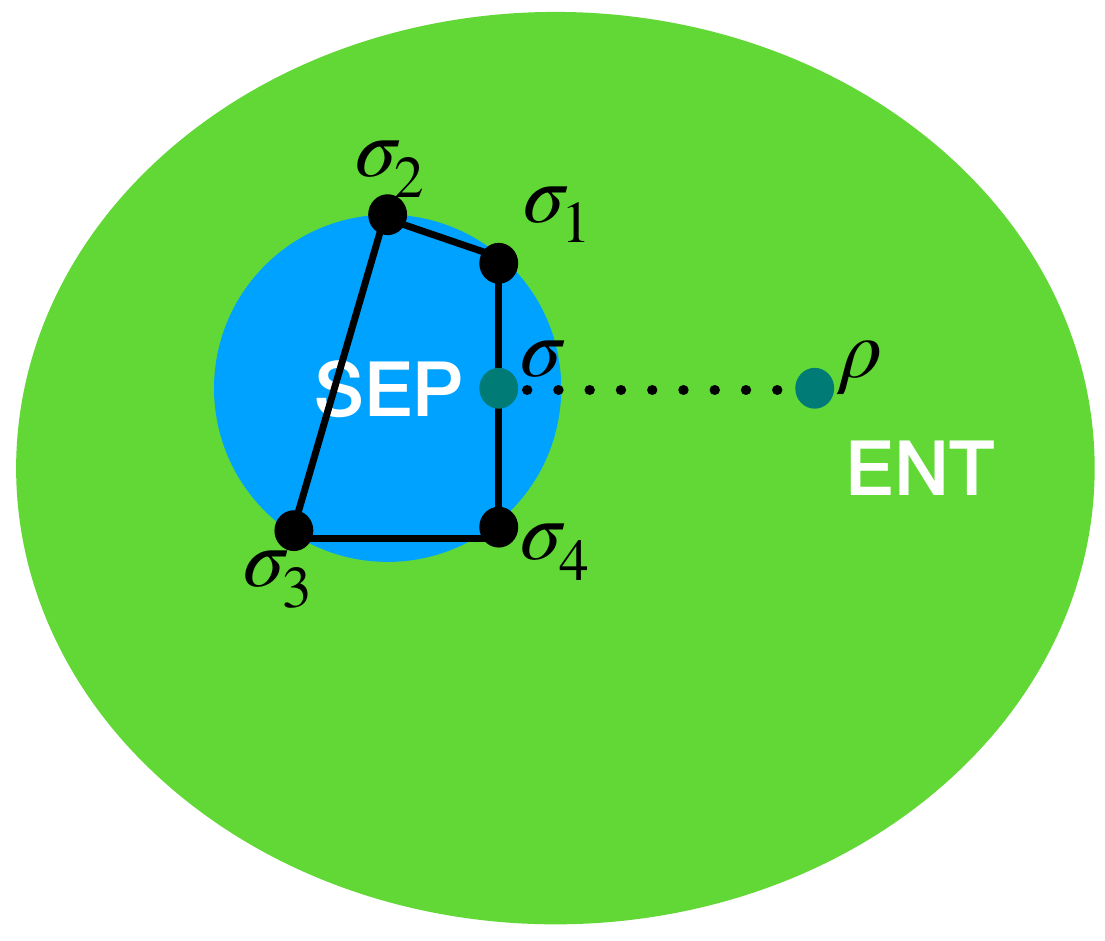}
\caption{Demonstration of the algorithm. We randomly sample $\sigma_1$, $\sigma_2$, ..., $\sigma_n$, and form a convex hull. Then use SDP to find $\sigma$. Based on the initial results, abandon $\sigma_1$s that are too far from the obtained $\sigma$. Sample more $\sigma_i$s, use SDP to find a new $\sigma$. Repeat until the results can no longer be improved.}
\label{fig:algo}
\end{figure}

Notice that the contribution of different $c_i$s are different. 
For example,
In~\cref{fig:algo}, we can see that  the final $\sigma$ is a linear combination of $\sigma_1$ and $\sigma_4$. States $\sigma_2$ and $\sigma_3$ has no contribution to $\sigma$. In this example, those points on the boundary without much contribution could be moved arbitrarily on the arc right to $\sigma_1$ and $\sigma_4$ without any effect on the final result, while if $\sigma_1$ and $\sigma_4$ move, the result would be significantly changed. Therefore, $\sigma_1$ and $\sigma_4$ could be labelled as \textit{useful} points, while $\sigma_2$ and $\sigma_3$ could be labelled as \textit{useless} points and can be discarded. 

However, the boundary of the set of separable states is way much more complicated than just a circle. Moreover, whether a point is useful does not remain unchanged, e.g. if we choose a point on the left of $\sigma_1$, then $\sigma_1$ becomes useless.
Since the extreme points are labelled as either \textit{useful} or \textit{useless}, a natural way is to use the method of active learning.

%Intuitively, we would want to build a classifier using supervised learning by constructing and training a deep neural network (DNN). However, there are to some obstacles. First, whether a point is useful or useless depends completely on what the given bipartite state $\rho$ is. Second, not a single one of the extreme points are labelled: we have countless extreme points, but none of them are classified. 
%Therefore, supervised learning does not work here. Active learning, a case of semi-supervised learning, is promising way of solving such a problem: there are a lot of data but few or even none of them are labelled.

%Active learning  does not require fully labelled data. Instead, 
%it aims to achieve high accuracy using as few labelled date as possible. There are three main scenarios for active learning: membership query synthesis, steam-base selective sampling and pool-based sampling. In membership query synthesis, the learner may request labelling for any data from the input space from a query. This setting not only works for classification problem, it also works for regression problem. 

Now we are ready to describe our algorithm based on active learning. First we could randomly generate some points $\sigma_i$ in the form of
\begin{equation}
\sigma_i=|i\rangle\langle i|\otimes |j\rangle\langle j|,
\end{equation}
where $\ket{i}$ and $\ket{j}$ are pure states randomly generated in $\mathcal{H}_A$ and $\mathcal{H}_B$. With these extreme points, we can construct a convex hull. On this convex hull, we perform SDP to minimize $E_R(\rho_{A})=\Tr(\rho\log\rho_{A}-\rho\log\sigma_{A'})$ and found a state 
\begin{equation}
\sigma=\sum_i c_i \sigma_i.
\end{equation}
Here we can set a threshold $\epsilon$, for those $\sigma_i$s, of which the corresponding $c_i\geq\epsilon$, we can say these points contribute significantly to the result and can be labelled as \textit{useful}. For those $\sigma_i$s of which the corresponding $c_i<\epsilon$, we can say these points has little contribution, hence \textit{useless} and to be discarded. Therefore, the SDP process works as the query or oracle in active learning. Take \cref{fig:algo} as an example, obviously, the solution is 
\begin{equation}
\sigma=c_1 \sigma_1 + c_2\sigma_2.
\end{equation}
$c_3$ and $c_4$ corresponding $\sigma_3$ and $\sigma_4$ are both 0. Therefore, $\sigma_1$ and $\sigma_2$ are labelled as \textit{useful}, while $\sigma_3$ and $\sigma_4$ are labelled as \textit{useless} and discarded. Then, if we want to obtain a better result, we can sample points near $\sigma_1$ and $\sigma_2$ and perform SDP again. Sampling near a state $|i\rangle$ can be done by
\begin{equation}
|i'\rangle=U|i\rangle,
\end{equation}
where the unitary operation
\begin{equation}
U=e^{-iH\delta},
\end{equation}
with $H$ being a randomly generated Hamiltonian and $\delta$ being a small number (time interval). Then we can perform the queries on the new samples again. By doing sampling and queries iteratively  we can get a relative result. 

\subsection{Effectiveness of active learning}

\begin{figure}[!htb]
\includegraphics[scale=0.4]{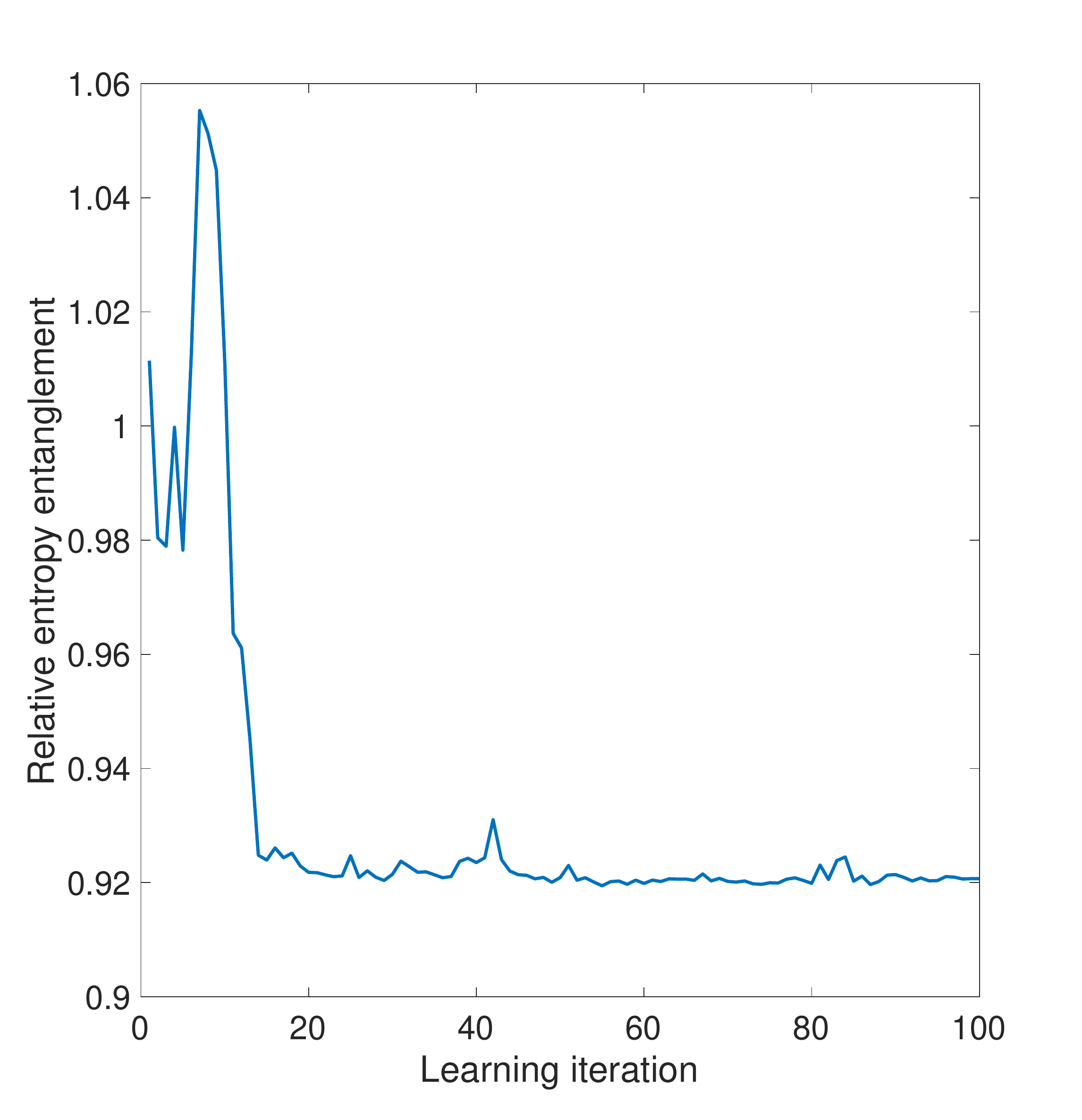}
\caption{Effect of active learning. The result becomes more accurate during the iterations.}
\label{fig:aliter}
\end{figure}

As an example, we show how our algorithm based on active learning approaches an upper bound for the relative entropy of entanglement $E_R(\rho)$.
In \cref{fig:aliter}, we demonstrate $E_R(\rho)$ during the iterations of active learning for state $\ket{00}+\ket{11}+\ket{12}$ with the dimension of $\mathcal{H}_A$ and $\mathcal{H}_B$ being $2$ and $3$, respectively. For such a state, the set of states with PPT is exactly the same as the set of separable states. The relative entropy of such a state obtained from PPT is $0.918$, which could be seen as the true value. From \cref{fig:aliter}, we can see that initially, just by sampling the points randomly, we obtain a result of $1.01$, which is quite larger than the ideal value. However, by using active learning, after $10$-$20$ iterations, the result get to close to the ideal value. By using active learning, we get a good approximation of the relative entropy of entanglement by a small number of queries and iterations, by avoiding the full labelling of all the extreme points, which is impossible.

\section{results}

In this section, we apply our method to various situations of interest.
We focus on bipartite systems of dimension $d_Ad_B$. We take three steps:
firstly we apply our method to calculate $E_R(\rho)$ for states
whose relative entropy of entanglement is analytically known, and compare 
our results with both the analytical value and the value given by the PPT method.
Then we take a step further to evaluate $E_R(\rho)$ for a set of example of
bound entangle states, where the PPT method cannot give useful information for 
$E_R(\rho)$, and demonstrate that our bound likely give a good approximation
for the real value of $E_R(\rho)$. Finally, we apply our method to randomly generated
states for different values of $d_Ad_B$, to show that 
the difference between the set of SEP and the set of PPT, for $d_Ad_B>6$.

In our numerical experiment of calculating $E_R$ based on active learning, we choose the number of $\sigma_i$'s to be $2000$ and the number of iterations for finding a better $\sigma$ to be $50$. We tested our algorithm on a workstation with Intel Xeon E5-2698 CPU and $256$ GB memory. The relationship between the average time cost for a random density matrix and the local dimensions is shown in Table \ref{tab:time}. According to our test, a majority of the portion of the time cost is due to the generation of random $\sigma_i$'s in each iteration.

\begin{table}[htbp]
	\centering
	\caption{Typical time cost for different system size}
	\begin{tabular}{c|r}
	\toprule[1pt]
		%\toprule  % 顶部线
		$d_A \otimes d_B$ & Time ($s$)\\ 
		\midrule[0.5pt]  % 中部线
		2$\otimes$2 &81.5 \\
		2$\otimes$3 &159.6\\
		2$\otimes$4 &338.8\\
		3$\otimes$3 &453.0\\
		3$\otimes$4 &1092.2\\
		4$\otimes$4 &2795.5 \\
		\bottomrule[1pt]  % 底部线
	\end{tabular}
	\label{tab:time}
\end{table}

\subsection{States with analytically known $E_R$}

%\subsubsection{Werner States}

To demonstrate the validity of our algorithm for 
calculating the upper bound of $E_R$, we first look at two kind of states whose 
$E_R$ are analytically known.

We start with the famous Werner states for $d\otimes d$ dimensional bipartite systems,
which are states invariant under any unitary transform with the form
$U\otimes U$~\cite{PhysRevA.40.4277}. 
There are several ways for parametrizing these states. Here, we use the following form to write a
Werner states as
\begin{equation}
\rho_W(\alpha)=\frac{1}{d^{2}-d\alpha}(I_{d^2}-\alpha F),
\end{equation} 
where $d$ is the dimension, and $F=\sum_{ij}|ij\rangle\langle ji|$ (which is actually a
swap operator). 

Whether a Werner state is entangled is determined by
the parameter $\alpha$, and can be given by the PPT criterion.
That is, if a Werner state has positive partial transpose, then it is separable, otherwise it 
is entangled. Therefore, when calculating $E_R$ for Werner states, optimization over either the 
set of SEP or PPT should give the same result.

It is known that for $2\otimes 2$ Werner states, the
state is separable if $\alpha \leq 1/2$ and entangled if
$\alpha > 1/2$. For $3\otimes 3$ Werner states, the state is separable if
$\alpha \leq 1/3$ and entangled if $\alpha > 1/3$~\cite{PhysRevA.40.4277}. 

Due to the symmetry, $E_R$
for Werner states can be
calculated analytically, which is given by~\cite{vollbrecht2001entanglement}
\begin{equation}
E_R(\rho_W)=e_R(\tr (\rho_W F)),
\end{equation}
where $F=\sum_{ij}|ij\rangle\langle ij|$, and $e_R$ is the function
\begin{equation}
e_R(f)=\log(2)-S(\frac{1+f}{2},\frac{1-f}{2}),
\end{equation}
with $S(p_1,p_2,...,p_n)=-\sum_k p_k\log p_k$.
In other words, for Werner state, the relative entropy of entanglement could be calculated by selecting $\sigma$ as the ``boundary" Werner state. That is, for $2\otimes 2$ Werner state, $\sigma=\rho_W(1/2)$ and for $3\otimes 3$ Werner state, $\sigma=\rho_W(1/3)$. 

\begin{figure*}[!ht]
\subfigure[ $2\otimes 2$ Werner states]{
\begin{minipage}{0.5\linewidth}
\includegraphics[width=0.98\linewidth]{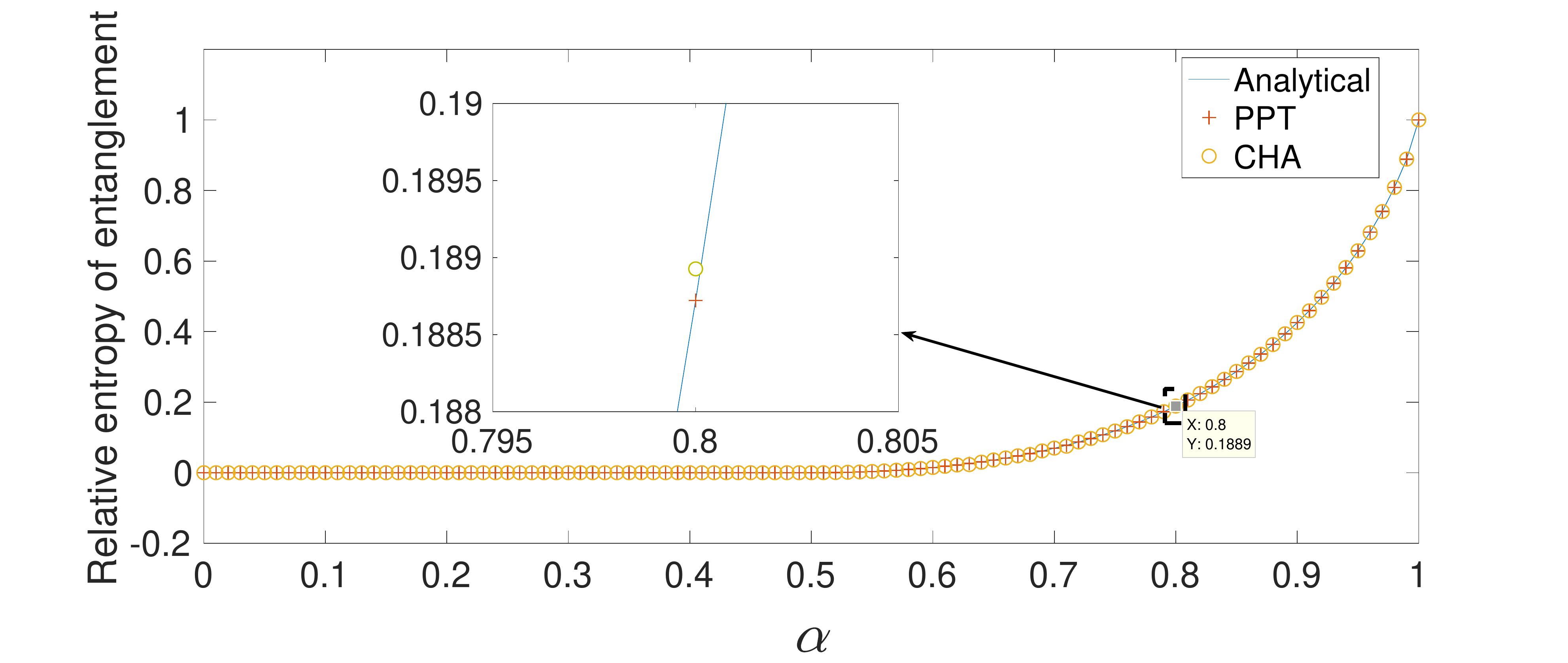}
\end{minipage}
}\subfigure[$3\otimes 3$ Werner states]{
\begin{minipage}{0.5\linewidth}
\includegraphics[width=0.98\linewidth]{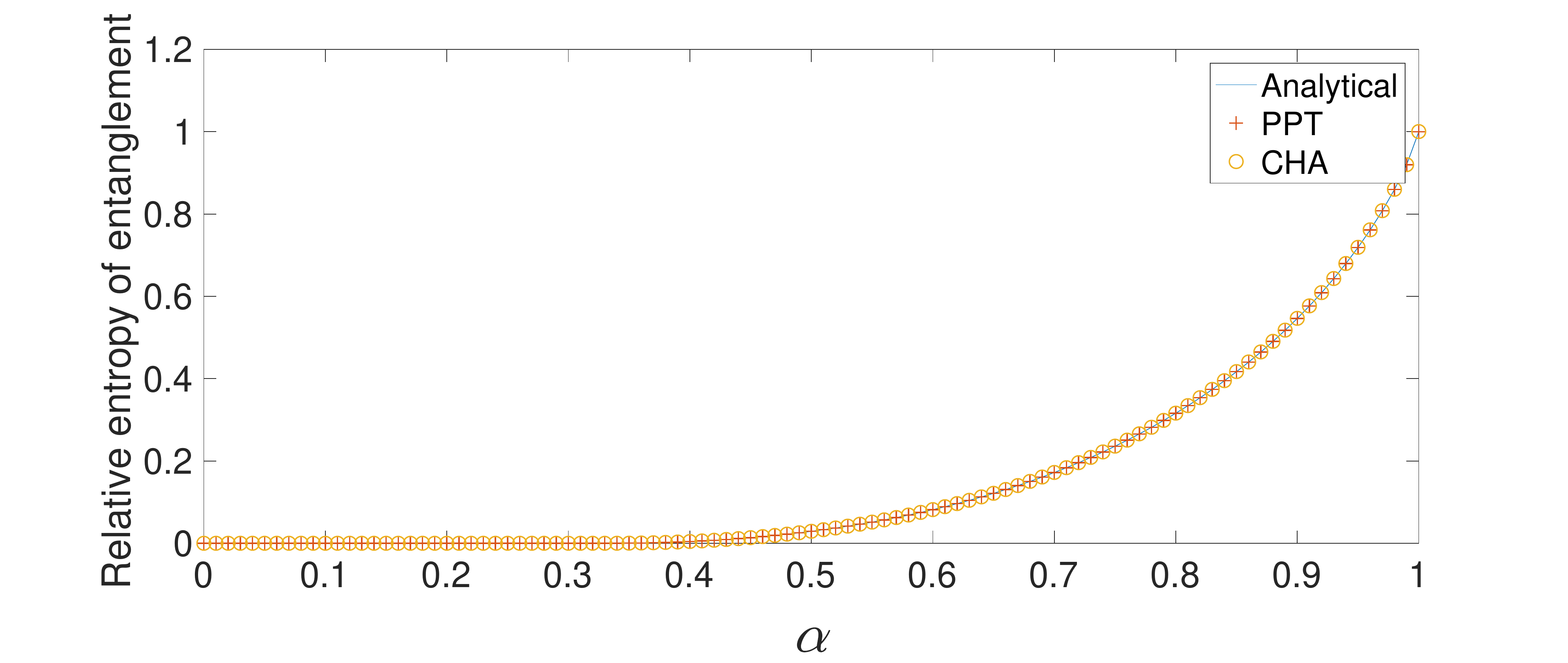}
\end{minipage}}
\caption{Results of the relative entropy of entanglement calculated using PPT and active learning for 2- and 3-dimension Werner states. The red crosses are the results of PPT and the orange circle are the results of active learning, with the blue line as the analytical results}
\label{fig:wer}
\end{figure*}

We performed the calculation of $E_R$
on $2\otimes 2$ and $3\otimes 3$ Werner states with three different methods: 
value from analytical formula,
method based on PPT optimization, and our method based on active learning. \
The results are shown in
\cref{fig:wer}.

As shown in \cref{fig:wer}, our approach using active learning as the approximation of 
$\mathcal D$ works perfectly for Werner states. The results of our approach, as well
as the previous method based on PPT, agree well with the analytical results for different
dimensions, see \cref{fig:wer}(a) and
\cref{fig:wer}(b).

Notice that, which is shown in the middle small figure in
\cref{fig:wer}(a), the results of our approach are slightly larger than
that of PPT, to a scale of 0.01\% of the actual analytical value. 
The reason is that our approach
approximates the set $\mathcal{D}$ from inside and in fact calculates a
upper bound. Therefore our results are always slightly larger than that is
given by the 
PPT method.

%\subsubsection{Isotropic states}

We then further look at isotropic states, which is a family of $d\otimes d$ dimension bipartite quantum
states that are invariant under any unitary transform with the form
$U\otimes U^{*}$ \cite{PhysRevA.59.4206}. 
The parametrized form of isotropic states is given by
\begin{equation}
\rho_i(\alpha)=\frac{1-\alpha}{d^2}I_{d^2}+\alpha |\psi_+\rangle\langle \psi_+|,
\end{equation} 
where $d$ is the dimension,
and $|\psi_+\rangle=\frac{1}{\sqrt{d}}\sum_j|jj\rangle$, which is actually a standard
maximally entangled state. 

Similar to the case of Werner states,
whether an isotropic state is entangled is
determined by the parameter $\alpha$, and can also be 
given by the PPT criterion. Therefore, when calculating $E_R$ for Werner states, optimization over either the 
set of SEP or PPT should give the same result.

It is known that for both $2\times 2$ and $3\times 3$ isotropic states, the state is separable if
$\alpha \leq 1/3$ and entangled is $\alpha > 1/3$. With good symmetry, the result
of relative entropy of entanglement of a certain isotropic state can
be calculated analytically~\cite{PhysRevA.60.179}.
For an isotropic state $\rho_i$, let $\hat{f}=\tr (\rho_i \hat{F})$ with $\hat{F}=\sum_{ij} |ii\rangle\langle jj|$. 
Then $E_R$ is given by 
\begin{eqnarray}
E_R(\rho_i)=\log (2)&-&(1-\frac{\hat{f}}{d})\log(d-1)\nonumber\\
&-&S(\frac{\hat{f}}{d},1-\frac{\hat{f}}{d}).
\end{eqnarray}
Similarly to Werner states, the calculation of relative entropy of entanglement of isotropic states could be carried out by choosing $\sigma$ as the boundary state, i.e. $\sigma=\rho_i(1/3)$.

\begin{figure*}[!ht]
\subfigure[ $2\otimes 2$ Isotropic states]{
\begin{minipage}{0.5\linewidth}
\includegraphics[width=0.98\linewidth]{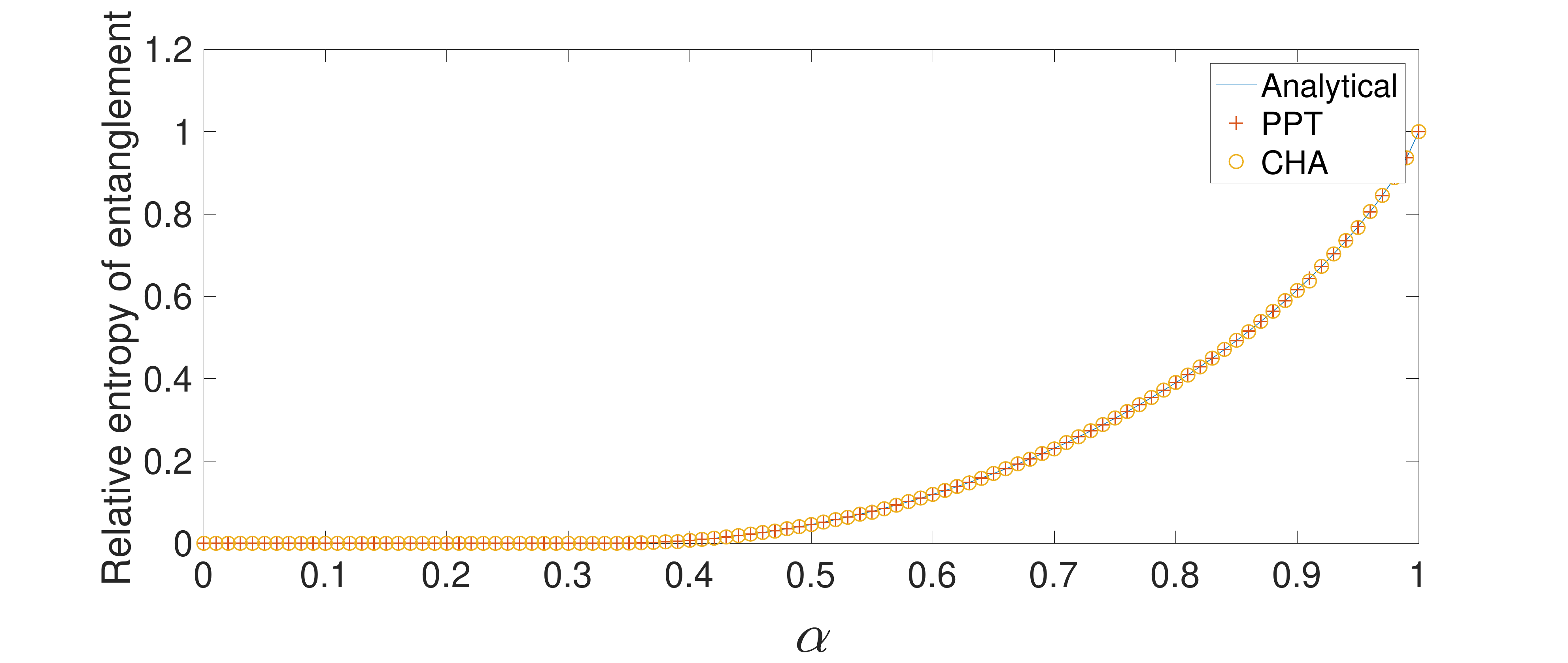}
\end{minipage}
}\subfigure[$3\otimes 3$ Isotropic states]{
\begin{minipage}{0.5\linewidth}
\includegraphics[width=0.98\linewidth]{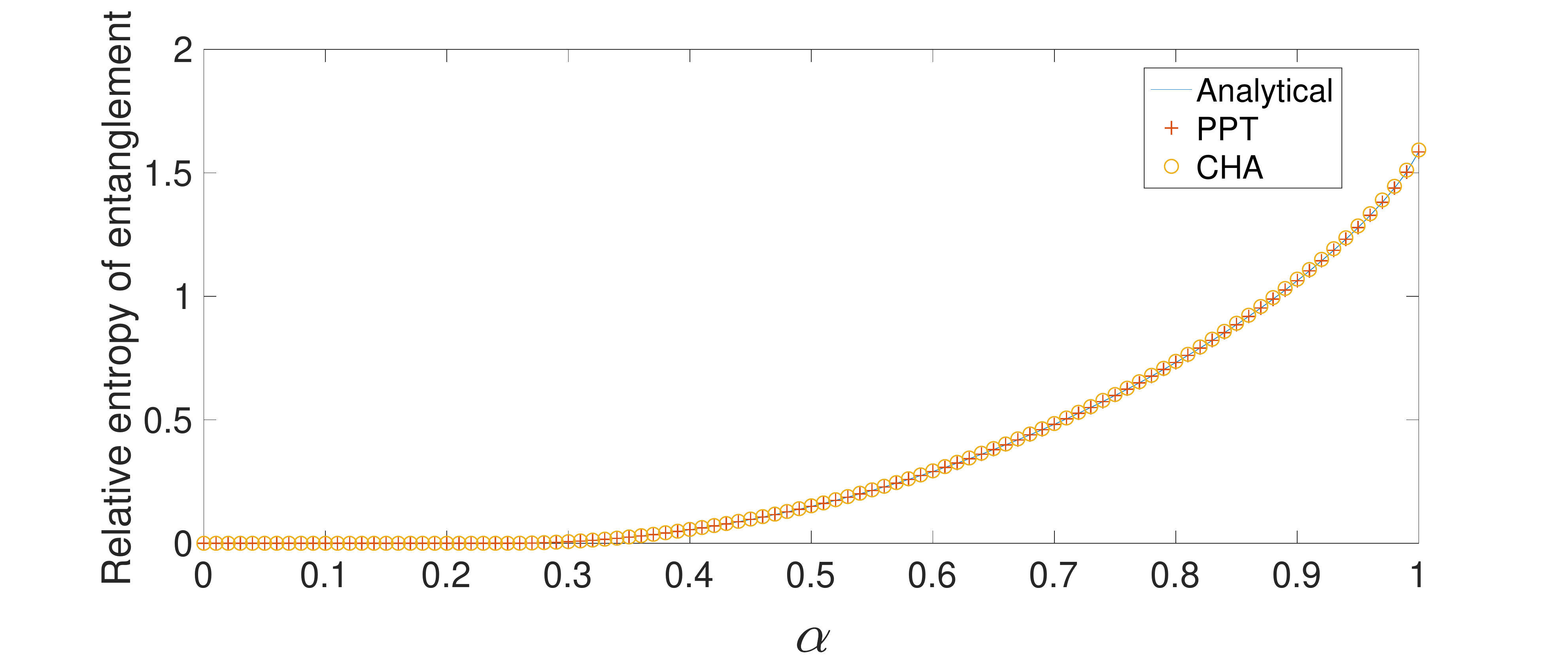}
\end{minipage}}
\caption{Results of the relative entropy of entanglement calculated using PPT and active learning for 2- and 3-dimension isotropic state. The red crosses are the results of PPT and the orange circle are the results of active learning, with the blue lines as the analytical results}
\label{fig:iso}
\end{figure*}

We performed the calculation of the relative entropy of entanglements
on $2\otimes 2$ and $3\otimes 3$ isotropic states with different methods: 
value from analytical formula,
method based on PPT optimization, and our method based on active learning.
The results are shown in
\cref{fig:iso}.

Similar to the results of Werner states, our approach using active learning as
the approximation of $\mathcal D$ works well for isotropic states. The
results of our approach, as well as SDP using PPT, agree well with the
analytical results for different dimensions, as shown in
\cref{fig:wer}(a) and \cref{fig:wer}(b). 

\subsection{Example for bound entangled states}

We now apply our algorithm to find upper bound of $E_R$ for 
bound entangled state, which are entangled states with 
positive partial transpose. In this case, the method based on PPT
will just give a value zero for a lower bound of $E_R$, which in fact
fails to give any information of $E_R$.
Our method based on active learning, instead, 
gives an upper bound for $E_R$, which in many cases can give results close to
the true value of $E_R$.

We consider the following example, where 
a set of two-qutrit pure states
$\{\ket{v_1},\dots,\ket{v_5}\}$ form the well known unextendible
product basis~\cite{bennett1999unextendible}: 
\begin{eqnarray}
\ket{v_1}&=&(\ket{00} -\ket{01} )/\sqrt{2},\nonumber\\
\ket{v_2}&=&(\ket{21} -\ket{22}) /\sqrt{2},\nonumber\\
\ket{v_3}&=&(\ket{02} -\ket{12}) /\sqrt{2},\nonumber\\
\ket{v_4}&=&(\ket{10} -\ket{20}) /\sqrt{2}, \nonumber\\
\ket{v_5}&=&(\ket{0}+\ket{1}+\ket{2})^{\otimes 2}/3. 
\end{eqnarray}

It is known that
\begin{equation}
\rho_{\text{tiles}} = (\mathbb{I}-\sum_{i=1}^{5}\ket{v_i}\bra{v_i} )/4
\end{equation}
is a bound entangled state. Therefore, the calculation
of $E_R$ for $\rho_{\text{tiles}}$ based on the PPT method returns the value zero.

Consider the following set of states with
parameter $\alpha$:
\begin{equation}
\rho_t(\alpha)=\alpha\rho_{\text{tiles}}+\frac{1-\alpha}{9}I.
\end{equation}
It is known that the critical
point for whether $\rho_t(\alpha)$ is entangled or not is
$\alpha\approx 0.8649$~\cite{sirui2018ent}.

\begin{figure}
\includegraphics[scale=0.22]{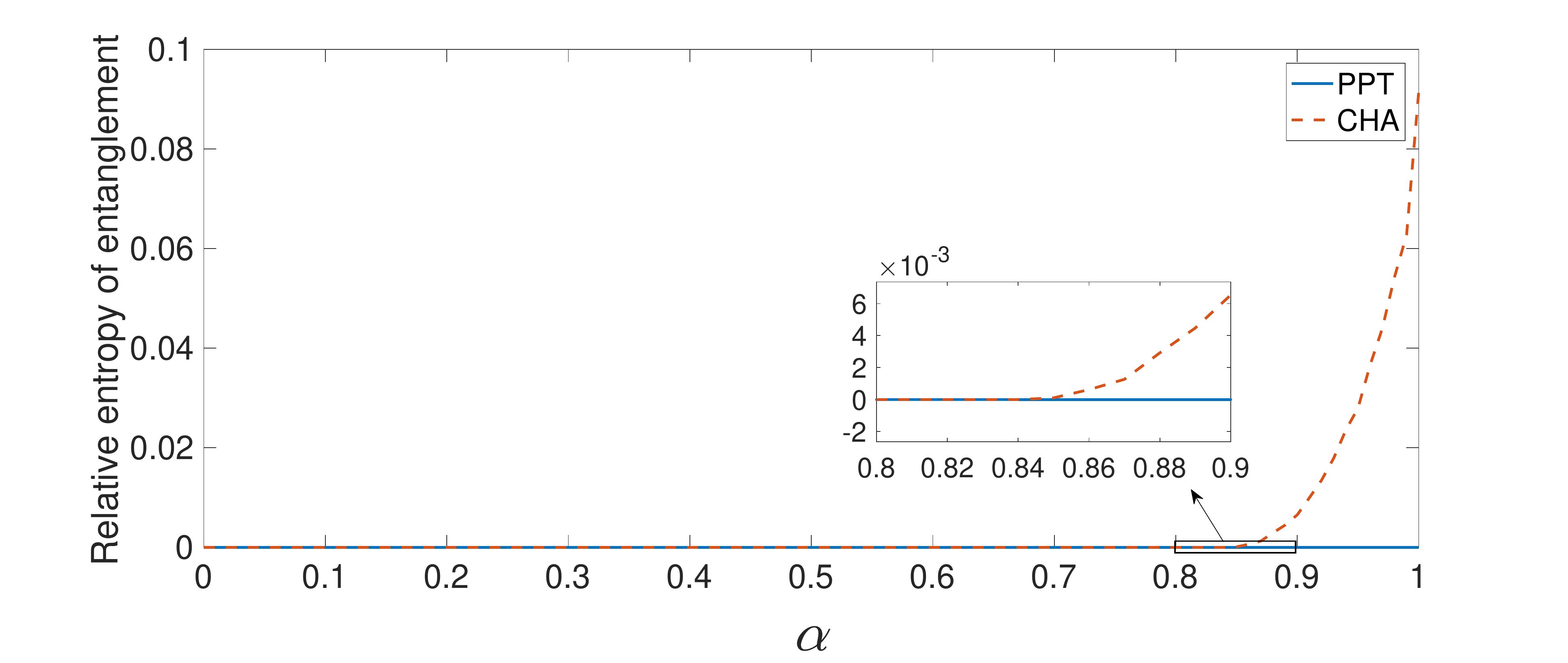}
\caption{Results of $E_R$ for $\rho_t(\alpha)$. Calculation using PPT will give the results of zero (blue line), while that using active learning will give a non-zero relative entropy of entanglement (orange line), which we think is closer to the actual one.}
\label{fig:pptes}
\end{figure}

We calculated $E_R$ for $\rho_t(\alpha)$ using based
on active learning,
and the results are shown in \cref{fig:pptes}. 
In the middle small figure of \cref{fig:pptes}, we see
that the relative entropy entanglement goes from zero to non-zero
value from around 0.86, which agree well with
critical value $\alpha\approx 0.8649$. 
These results demonstrate the effectiveness 
of the active learning approach for finding upper bounds
of $E_R$ that is close to the true values.

\subsection{Random States}

We apply our algorithm to find upper bounds
of $E_R$ for bipartite system of 
various dimensions of $d_A,d_B$,
and compare with the lower bounds
based on the PPT method.

\begin{figure*}[!htb]
\subfigure[ $2\otimes 2$]{
\begin{minipage}{0.45\linewidth}
\includegraphics[width=0.98\linewidth]{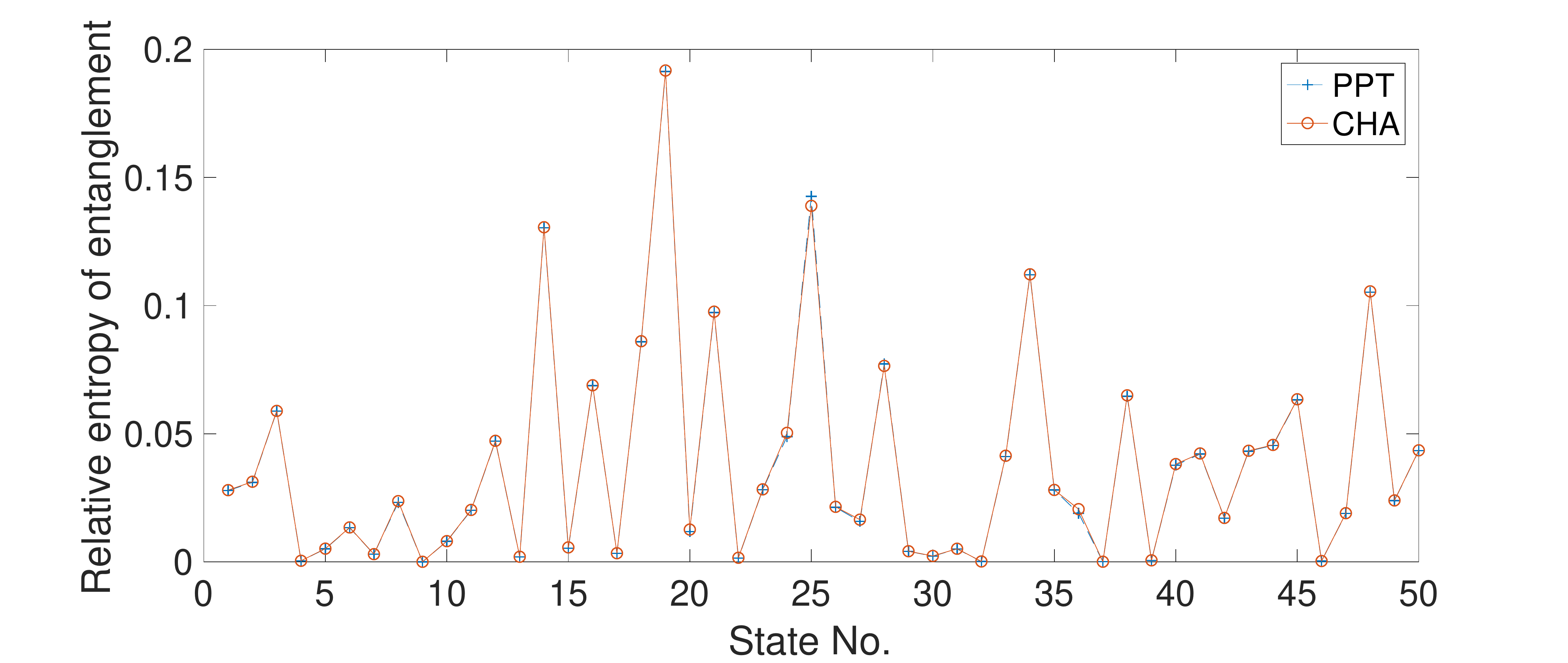}
\end{minipage}
}\subfigure[$2\otimes 3$]{
\begin{minipage}{0.45\linewidth}
\includegraphics[width=0.98\linewidth]{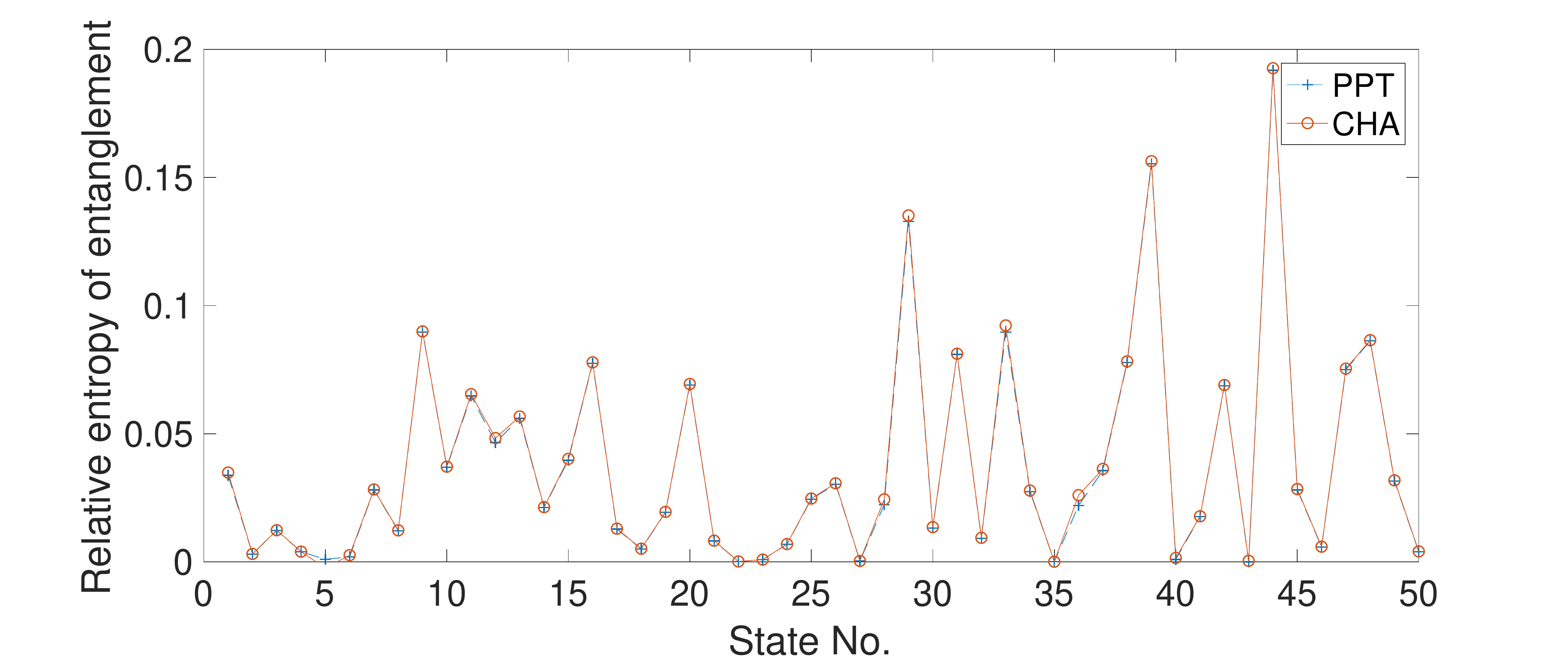}
\end{minipage}}\\

\subfigure[$2\otimes 4$]{
\begin{minipage}{0.45\linewidth}
\includegraphics[width=0.98\linewidth]{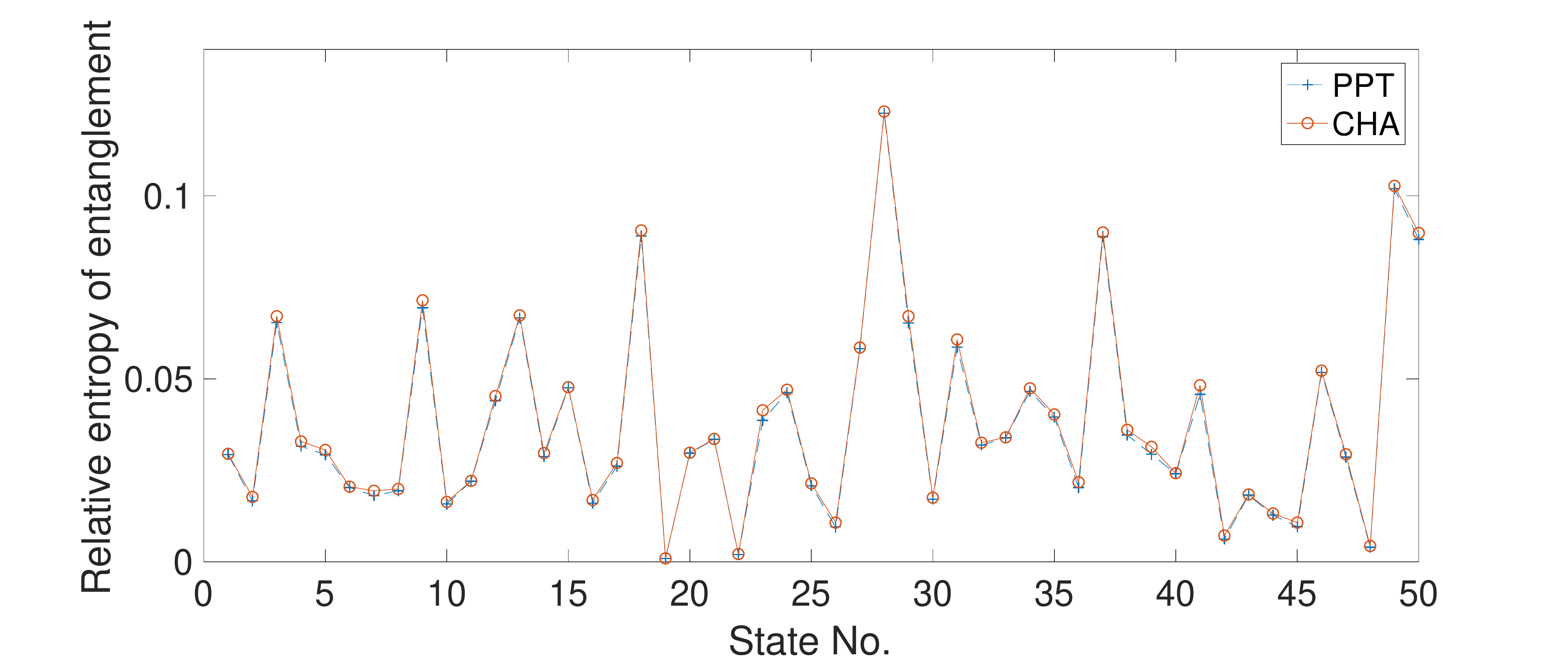}
\end{minipage}
}\subfigure[$3\otimes 3$]{
\begin{minipage}{0.45\linewidth}
\includegraphics[width=0.98\linewidth]{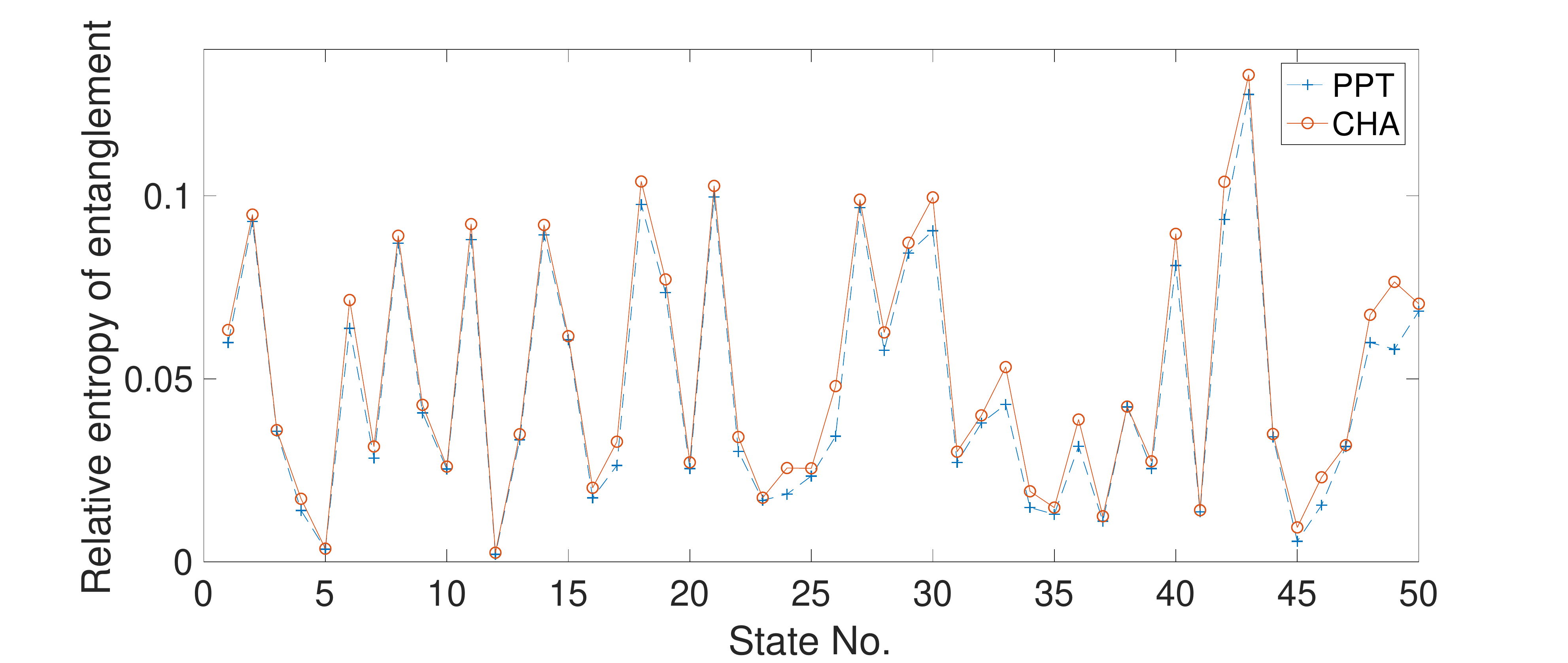}
\end{minipage}
}\\

\subfigure[$3\otimes 4$]{
\begin{minipage}{0.45\linewidth}
\includegraphics[width=0.98\linewidth]{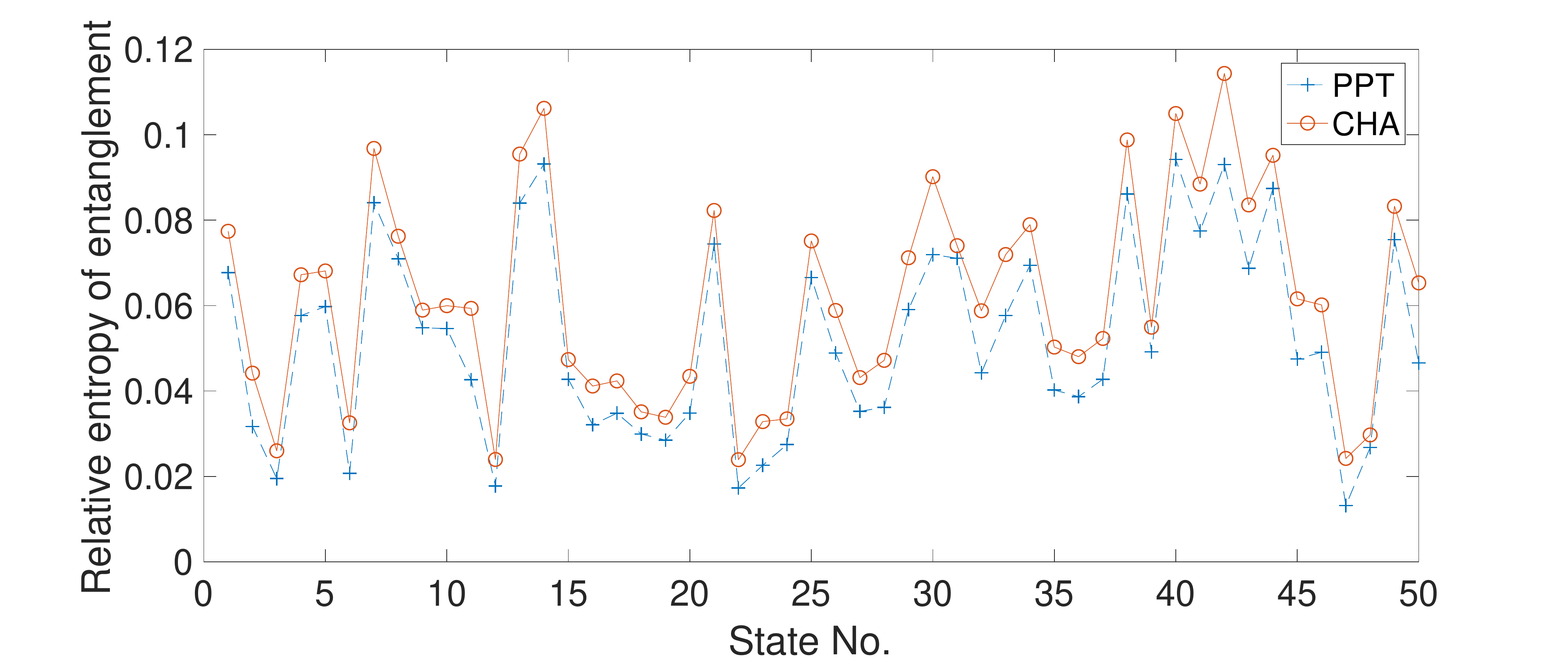}
\end{minipage}
}\subfigure[$4\otimes 4$]{
\begin{minipage}{0.45\linewidth}
\includegraphics[width=0.98\linewidth]{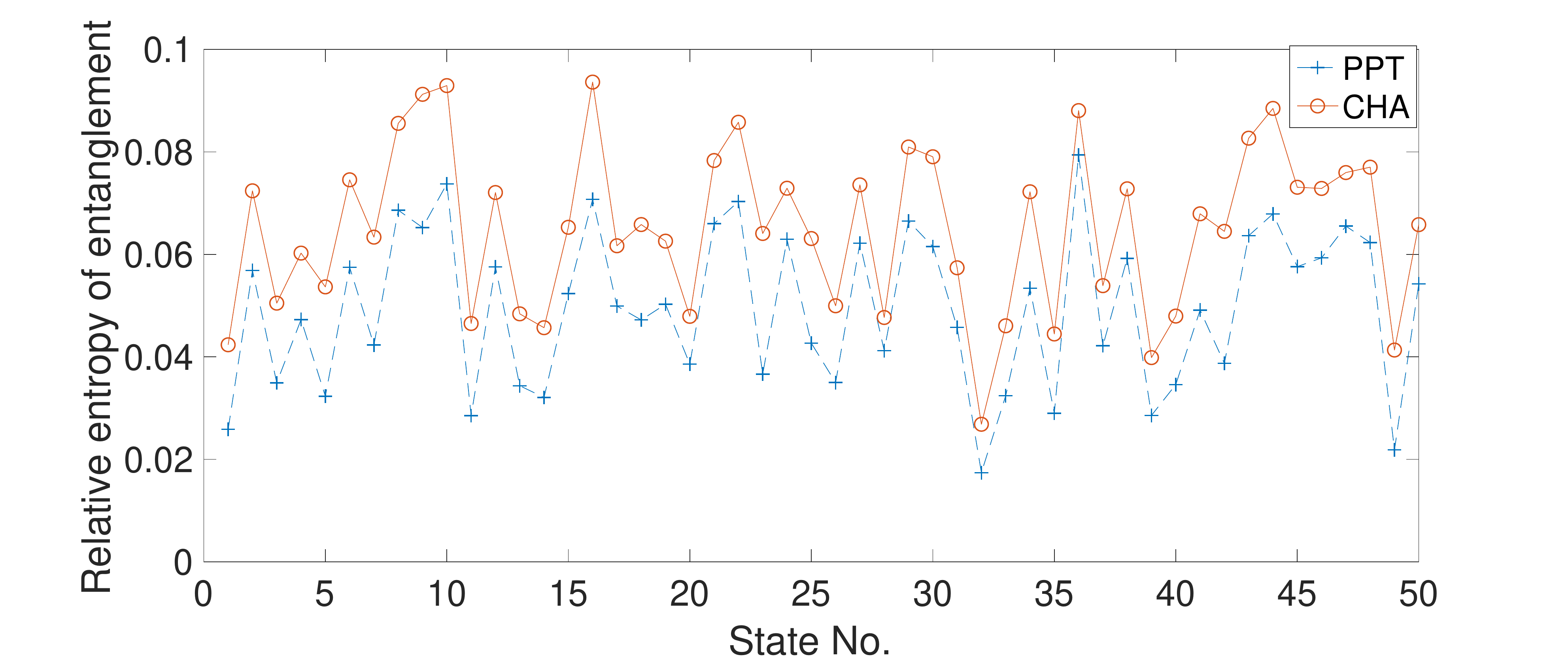}
\end{minipage}
}
\caption{Results for random entangled states.}
\label{fig:rand}
\end{figure*}

We consider 4 cases, with dimensions
$2\otimes 2$, $2\otimes 3$, $2\otimes 4$, and $3\otimes 3$. 
For each case, we randomly
generated $50$ entangled states using a
similar method discussed in~\cite{sirui2018ent}). We then calculate 
the upper bound of $E_R$ based on the active learning method,
and the lower bound of $E_R$ based on the PPT method.
All the results are shown in \cref{fig:rand}.

Since for 
$d_Ad_B\leq 6$, the set of SEP is the same as the set of PPT, 
the upper bound and lower bound should be very close to each other. 
This is indeed what we see for the case of $2\otimes 2$ and $2\otimes 3$,
as shown in \cref{fig:rand}(a) and \cref{fig:rand}(b).
%we can see that the
%results of CHA and PPT agree well with each other within numeric
%error. 
For most cases, the two bound coincide to 
give the true value of $E_R$.
For some states, the results of active learning is slightly larger than
that of PPT. 
The results clearly meet the predictions of our
theoretical analysis, demonstrating the effectiveness of our method.

For $d_Ad_B> 6$, the set of SEP is strictly smaller than 
that of PPT. For random states, one would expect gap
between the upper bounds given by active learning and the lower bounds
given by PPT. This indeed what we see in the cases of 
$2\otimes 4$ and
$3\otimes 3$,
as shown in \cref{fig:rand}(c), \cref{fig:rand}(d), \cref{fig:rand}(e), and \cref{fig:rand}(f).

In the cases of $2\otimes 4$ and $3 \otimes 3$,
we see that for  most states,
the results of active learning is still very close to that of PPT. However, for some states, the
results of active learning is significantly larger than that of PPT. And the
gap is in fact larger in the case of $3 \otimes 3$ compared to the 
case of $2\otimes 4$. In the cases of 
$3\otimes 4$ and $4\otimes 4$, for almost all states sampled, 
the upper bounds of $E_R$ 
calculated from active learning are larger than that of PPT.
These observations agree with the previous observation
on the volume of SEP compared to PPT~\cite{zyczkowskiVolumeI,zyczkowskiVolumeII}. With the values of 
upper bound of $E_R$, our results also provide new tools to study such
volume and deepen our understanding of the difference between 
SEP and PPT.

\section{Discussion}

In this work, we propose a reliable method for calculating upper bounds for relative entropy of entanglement, based
on active learning to generate an approximation for the set of separable states. We apply our method to calculate the upper bounds of the measure for the composite system of various sizes, and compare to the previous known lower bounds, obtaining promising results. Since the active learning approach is a powerful general idea to approximate convex set from inside, our method is naturally generalizable for obtaining upper bounds for any quantity of interest with optimization over convex set, especially in the case that extreme points of these sets are relatively easy to sample. We hope that our work adds new understanding on the structure of separable states, provides further information on the difference between the set of separable states and PPT states in various dimensions, and sheds light on the calculation of relevant quantities based on optimization over convex sets.

\section*{Acknowledgement}

We thank Sirui Lu for helpful discussions. S.-Y.H is supported by National Natural Science Foundation of China under Grant No. 11847154. D.L.Zhou is supported from NSF of China (Grant No.11775300), the National Key Research and Development Program of China (2016YFA0300603), and the Strategic Priority Research Program of Chinese Academy of Sciences No. XDB28000000.

%\appendix

%\section{Parameter and performance of the algorithm}

\bibliography{ree}

\begin{thebibliography}{35}
\expandafter\ifx\csname natexlab\endcsname\relax\def\natexlab#1{#1}\fi
\expandafter\ifx\csname bibnamefont\endcsname\relax
  \def\bibnamefont#1{#1}\fi
\expandafter\ifx\csname bibfnamefont\endcsname\relax
  \def\bibfnamefont#1{#1}\fi
\expandafter\ifx\csname citenamefont\endcsname\relax
  \def\citenamefont#1{#1}\fi
\expandafter\ifx\csname url\endcsname\relax
  \def\url#1{\texttt{#1}}\fi
\expandafter\ifx\csname urlprefix\endcsname\relax\def\urlprefix{URL }\fi
\providecommand{\bibinfo}[2]{#2}
\providecommand{\eprint}[2][]{\url{#2}}

\bibitem[{\citenamefont{Horodecki et~al.}(2009)\citenamefont{Horodecki,
  Horodecki, Horodecki, and Horodecki}}]{horodecki2009quantum}
\bibinfo{author}{\bibfnamefont{R.}~\bibnamefont{Horodecki}},
  \bibinfo{author}{\bibfnamefont{P.}~\bibnamefont{Horodecki}},
  \bibinfo{author}{\bibfnamefont{M.}~\bibnamefont{Horodecki}},
  \bibnamefont{and}
  \bibinfo{author}{\bibfnamefont{K.}~\bibnamefont{Horodecki}},
  \bibinfo{journal}{Reviews of modern physics} \textbf{\bibinfo{volume}{81}},
  \bibinfo{pages}{865} (\bibinfo{year}{2009}).

\bibitem[{\citenamefont{Lloyd}(1997)}]{lloyd1997capacity}
\bibinfo{author}{\bibfnamefont{S.}~\bibnamefont{Lloyd}},
  \bibinfo{journal}{Physical Review A} \textbf{\bibinfo{volume}{55}},
  \bibinfo{pages}{1613} (\bibinfo{year}{1997}).

\bibitem[{\citenamefont{Farhi et~al.}(2001)\citenamefont{Farhi, Goldstone,
  Gutmann, Lapan, Lundgren, and Preda}}]{farhi2001quantum}
\bibinfo{author}{\bibfnamefont{E.}~\bibnamefont{Farhi}},
  \bibinfo{author}{\bibfnamefont{J.}~\bibnamefont{Goldstone}},
  \bibinfo{author}{\bibfnamefont{S.}~\bibnamefont{Gutmann}},
  \bibinfo{author}{\bibfnamefont{J.}~\bibnamefont{Lapan}},
  \bibinfo{author}{\bibfnamefont{A.}~\bibnamefont{Lundgren}}, \bibnamefont{and}
  \bibinfo{author}{\bibfnamefont{D.}~\bibnamefont{Preda}},
  \bibinfo{journal}{Science} \textbf{\bibinfo{volume}{292}},
  \bibinfo{pages}{472} (\bibinfo{year}{2001}).

\bibitem[{\citenamefont{Bennett et~al.}(1996)\citenamefont{Bennett, DiVincenzo,
  Smolin, and Wootters}}]{bennett1996mixed}
\bibinfo{author}{\bibfnamefont{C.~H.} \bibnamefont{Bennett}},
  \bibinfo{author}{\bibfnamefont{D.~P.} \bibnamefont{DiVincenzo}},
  \bibinfo{author}{\bibfnamefont{J.~A.} \bibnamefont{Smolin}},
  \bibnamefont{and} \bibinfo{author}{\bibfnamefont{W.~K.}
  \bibnamefont{Wootters}}, \bibinfo{journal}{Physical Review A}
  \textbf{\bibinfo{volume}{54}}, \bibinfo{pages}{3824} (\bibinfo{year}{1996}).

\bibitem[{\citenamefont{Huver et~al.}(2008)\citenamefont{Huver, Wildfeuer, and
  Dowling}}]{huver2008entangled}
\bibinfo{author}{\bibfnamefont{S.~D.} \bibnamefont{Huver}},
  \bibinfo{author}{\bibfnamefont{C.~F.} \bibnamefont{Wildfeuer}},
  \bibnamefont{and} \bibinfo{author}{\bibfnamefont{J.~P.}
  \bibnamefont{Dowling}}, \bibinfo{journal}{Physical Review A}
  \textbf{\bibinfo{volume}{78}}, \bibinfo{pages}{063828}
  (\bibinfo{year}{2008}).

\bibitem[{\citenamefont{Zeng et~al.}(2019)\citenamefont{Zeng, Chen, Zhou, and
  Wen}}]{zeng2019quantum}
\bibinfo{author}{\bibfnamefont{B.}~\bibnamefont{Zeng}},
  \bibinfo{author}{\bibfnamefont{X.}~\bibnamefont{Chen}},
  \bibinfo{author}{\bibfnamefont{D.-L.} \bibnamefont{Zhou}}, \bibnamefont{and}
  \bibinfo{author}{\bibfnamefont{X.-G.} \bibnamefont{Wen}},
  \emph{\bibinfo{title}{Quantum Information Meets Quantum Matter: From Quantum
  Entanglement to Topological Phases of Many-Body Systems}}
  (\bibinfo{publisher}{Springer}, \bibinfo{year}{2019}).

\bibitem[{\citenamefont{Almheiri et~al.}(2015)\citenamefont{Almheiri, Dong, and
  Harlow}}]{almheiri2015bulk}
\bibinfo{author}{\bibfnamefont{A.}~\bibnamefont{Almheiri}},
  \bibinfo{author}{\bibfnamefont{X.}~\bibnamefont{Dong}}, \bibnamefont{and}
  \bibinfo{author}{\bibfnamefont{D.}~\bibnamefont{Harlow}},
  \bibinfo{journal}{Journal of High Energy Physics}
  \textbf{\bibinfo{volume}{2015}}, \bibinfo{pages}{163} (\bibinfo{year}{2015}).

\bibitem[{\citenamefont{G{\"u}hne and T{\'o}th}(2009)}]{guhne2009entanglement}
\bibinfo{author}{\bibfnamefont{O.}~\bibnamefont{G{\"u}hne}} \bibnamefont{and}
  \bibinfo{author}{\bibfnamefont{G.}~\bibnamefont{T{\'o}th}},
  \bibinfo{journal}{Physics Reports} \textbf{\bibinfo{volume}{474}},
  \bibinfo{pages}{1} (\bibinfo{year}{2009}).

\bibitem[{\citenamefont{Vedral et~al.}(1997)\citenamefont{Vedral, Plenio,
  Rippin, and Knight}}]{vedral1997quantifying}
\bibinfo{author}{\bibfnamefont{V.}~\bibnamefont{Vedral}},
  \bibinfo{author}{\bibfnamefont{M.~B.} \bibnamefont{Plenio}},
  \bibinfo{author}{\bibfnamefont{M.~A.} \bibnamefont{Rippin}},
  \bibnamefont{and} \bibinfo{author}{\bibfnamefont{P.~L.}
  \bibnamefont{Knight}}, \bibinfo{journal}{Physical Review Letters}
  \textbf{\bibinfo{volume}{78}}, \bibinfo{pages}{2275} (\bibinfo{year}{1997}).

\bibitem[{\citenamefont{Vedral and Plenio}(1998)}]{vedral1998entanglement}
\bibinfo{author}{\bibfnamefont{V.}~\bibnamefont{Vedral}} \bibnamefont{and}
  \bibinfo{author}{\bibfnamefont{M.~B.} \bibnamefont{Plenio}},
  \bibinfo{journal}{Physical Review A} \textbf{\bibinfo{volume}{57}},
  \bibinfo{pages}{1619} (\bibinfo{year}{1998}).

\bibitem[{\citenamefont{Vedral}(2002)}]{vedral2002role}
\bibinfo{author}{\bibfnamefont{V.}~\bibnamefont{Vedral}},
  \bibinfo{journal}{Reviews of Modern Physics} \textbf{\bibinfo{volume}{74}},
  \bibinfo{pages}{197} (\bibinfo{year}{2002}).

\bibitem[{\citenamefont{Horodecki et~al.}(2000)\citenamefont{Horodecki,
  Horodecki, and Horodecki}}]{horodecki2000limits}
\bibinfo{author}{\bibfnamefont{M.}~\bibnamefont{Horodecki}},
  \bibinfo{author}{\bibfnamefont{P.}~\bibnamefont{Horodecki}},
  \bibnamefont{and}
  \bibinfo{author}{\bibfnamefont{R.}~\bibnamefont{Horodecki}},
  \bibinfo{journal}{Physical Review Letters} \textbf{\bibinfo{volume}{84}},
  \bibinfo{pages}{2014} (\bibinfo{year}{2000}).

\bibitem[{\citenamefont{Rains}(2001)}]{rains2001semidefinite}
\bibinfo{author}{\bibfnamefont{E.~M.} \bibnamefont{Rains}},
  \bibinfo{journal}{IEEE Transactions on Information Theory}
  \textbf{\bibinfo{volume}{47}}, \bibinfo{pages}{2921} (\bibinfo{year}{2001}).

\bibitem[{\citenamefont{Henderson and Vedral}(2000)}]{henderson2000information}
\bibinfo{author}{\bibfnamefont{L.}~\bibnamefont{Henderson}} \bibnamefont{and}
  \bibinfo{author}{\bibfnamefont{V.}~\bibnamefont{Vedral}},
  \bibinfo{journal}{Physical review letters} \textbf{\bibinfo{volume}{84}},
  \bibinfo{pages}{2263} (\bibinfo{year}{2000}).

\bibitem[{\citenamefont{Girard et~al.}(2014)\citenamefont{Girard, Gour, and
  Friedland}}]{girard2014convex}
\bibinfo{author}{\bibfnamefont{M.~W.} \bibnamefont{Girard}},
  \bibinfo{author}{\bibfnamefont{G.}~\bibnamefont{Gour}}, \bibnamefont{and}
  \bibinfo{author}{\bibfnamefont{S.}~\bibnamefont{Friedland}},
  \bibinfo{journal}{Journal of Physics A: Mathematical and Theoretical}
  \textbf{\bibinfo{volume}{47}}, \bibinfo{pages}{505302}
  (\bibinfo{year}{2014}).

\bibitem[{\citenamefont{Horodecki}(1997)}]{horodecki1997separability}
\bibinfo{author}{\bibfnamefont{P.}~\bibnamefont{Horodecki}},
  \bibinfo{journal}{Physics Letters A} \textbf{\bibinfo{volume}{232}},
  \bibinfo{pages}{333} (\bibinfo{year}{1997}).

\bibitem[{\citenamefont{Fawzi and Fawzi}(2018)}]{Fawzi_2018}
\bibinfo{author}{\bibfnamefont{H.}~\bibnamefont{Fawzi}} \bibnamefont{and}
  \bibinfo{author}{\bibfnamefont{O.}~\bibnamefont{Fawzi}},
  \bibinfo{journal}{Journal of Physics A: Mathematical and Theoretical}
  \textbf{\bibinfo{volume}{51}}, \bibinfo{pages}{154003}
  (\bibinfo{year}{2018}),
  \urlprefix\url{https://doi.org/10.1088%2F1751-8121%2Faab285}.

\bibitem[{\citenamefont{Miranowicz and
  Grudka}(2004)}]{miranowicz2004comparative}
\bibinfo{author}{\bibfnamefont{A.}~\bibnamefont{Miranowicz}} \bibnamefont{and}
  \bibinfo{author}{\bibfnamefont{A.}~\bibnamefont{Grudka}},
  \bibinfo{journal}{Journal of Optics B: Quantum and Semiclassical Optics}
  \textbf{\bibinfo{volume}{6}}, \bibinfo{pages}{542} (\bibinfo{year}{2004}).

\bibitem[{\citenamefont{Lu et~al.}(2018)\citenamefont{Lu, Huang, Li, Li, Chen,
  Lu, Ji, Shen, Zhou, and Zeng}}]{sirui2018ent}
\bibinfo{author}{\bibfnamefont{S.}~\bibnamefont{Lu}},
  \bibinfo{author}{\bibfnamefont{S.}~\bibnamefont{Huang}},
  \bibinfo{author}{\bibfnamefont{K.}~\bibnamefont{Li}},
  \bibinfo{author}{\bibfnamefont{J.}~\bibnamefont{Li}},
  \bibinfo{author}{\bibfnamefont{J.}~\bibnamefont{Chen}},
  \bibinfo{author}{\bibfnamefont{D.}~\bibnamefont{Lu}},
  \bibinfo{author}{\bibfnamefont{Z.}~\bibnamefont{Ji}},
  \bibinfo{author}{\bibfnamefont{Y.}~\bibnamefont{Shen}},
  \bibinfo{author}{\bibfnamefont{D.}~\bibnamefont{Zhou}}, \bibnamefont{and}
  \bibinfo{author}{\bibfnamefont{B.}~\bibnamefont{Zeng}},
  \bibinfo{journal}{Phys. Rev. A} \textbf{\bibinfo{volume}{98}},
  \bibinfo{pages}{012315} (\bibinfo{year}{2018}),
  \urlprefix\url{https://link.aps.org/doi/10.1103/PhysRevA.98.012315}.

\bibitem[{\citenamefont{Castro and Nowak}(2008)}]{castro2008minimax}
\bibinfo{author}{\bibfnamefont{R.~M.} \bibnamefont{Castro}} \bibnamefont{and}
  \bibinfo{author}{\bibfnamefont{R.~D.} \bibnamefont{Nowak}},
  \bibinfo{journal}{IEEE Transactions on Information Theory}
  \textbf{\bibinfo{volume}{54}}, \bibinfo{pages}{2339} (\bibinfo{year}{2008}).

\bibitem[{\citenamefont{Freund et~al.}(1997)\citenamefont{Freund, Seung,
  Shamir, and Tishby}}]{freund1997selective}
\bibinfo{author}{\bibfnamefont{Y.}~\bibnamefont{Freund}},
  \bibinfo{author}{\bibfnamefont{H.~S.} \bibnamefont{Seung}},
  \bibinfo{author}{\bibfnamefont{E.}~\bibnamefont{Shamir}}, \bibnamefont{and}
  \bibinfo{author}{\bibfnamefont{N.}~\bibnamefont{Tishby}},
  \bibinfo{journal}{Machine learning} \textbf{\bibinfo{volume}{28}},
  \bibinfo{pages}{133} (\bibinfo{year}{1997}).

\bibitem[{\citenamefont{Balcan et~al.}(2009)\citenamefont{Balcan, Beygelzimer,
  and Langford}}]{balcan2009agnostic}
\bibinfo{author}{\bibfnamefont{M.-F.} \bibnamefont{Balcan}},
  \bibinfo{author}{\bibfnamefont{A.}~\bibnamefont{Beygelzimer}},
  \bibnamefont{and} \bibinfo{author}{\bibfnamefont{J.}~\bibnamefont{Langford}},
  \bibinfo{journal}{Journal of Computer and System Sciences}
  \textbf{\bibinfo{volume}{75}}, \bibinfo{pages}{78} (\bibinfo{year}{2009}).

\bibitem[{\citenamefont{Gurvits}(2003)}]{gurvits2003classical}
\bibinfo{author}{\bibfnamefont{L.}~\bibnamefont{Gurvits}}, in
  \emph{\bibinfo{booktitle}{Proceedings of the thirty-fifth annual ACM
  symposium on Theory of computing}} (\bibinfo{organization}{ACM},
  \bibinfo{year}{2003}), pp. \bibinfo{pages}{10--19}.

\bibitem[{\citenamefont{Slater}(2019)}]{slater2019}
\bibinfo{author}{\bibfnamefont{P.~B.} \bibnamefont{Slater}},
  \bibinfo{journal}{arXiv preprint quant-ph/1905.09228}
  (\bibinfo{year}{2019}).

\bibitem[{\citenamefont{Johnston}(2014)}]{johnston2014detection}
\bibinfo{author}{\bibfnamefont{N.}~\bibnamefont{Johnston}},
  \emph{\bibinfo{title}{Entanglement detection}} (\bibinfo{year}{2014}).

\bibitem[{\citenamefont{Fawzi et~al.}(2018)\citenamefont{Fawzi, Saunderson, and
  Parrilo}}]{cvxquad}
\bibinfo{author}{\bibfnamefont{H.}~\bibnamefont{Fawzi}},
  \bibinfo{author}{\bibfnamefont{J.}~\bibnamefont{Saunderson}},
  \bibnamefont{and} \bibinfo{author}{\bibfnamefont{P.~A.}
  \bibnamefont{Parrilo}}, \bibinfo{journal}{Foundations of Computational
  Mathematics}  (\bibinfo{year}{2018}), \bibinfo{note}{package cvxquad at
  \url{https://github.com/hfawzi/cvxquad}}.

\bibitem[{\citenamefont{Grant and Boyd}(2014)}]{cvx}
\bibinfo{author}{\bibfnamefont{M.}~\bibnamefont{Grant}} \bibnamefont{and}
  \bibinfo{author}{\bibfnamefont{S.}~\bibnamefont{Boyd}},
  \emph{\bibinfo{title}{{CVX}: Matlab software for disciplined convex
  programming, version 2.1}}, \bibinfo{howpublished}{\url{http://cvxr.com/cvx}}
  (\bibinfo{year}{2014}).

\bibitem[{\citenamefont{Grant and Boyd}(2008)}]{gb08}
\bibinfo{author}{\bibfnamefont{M.}~\bibnamefont{Grant}} \bibnamefont{and}
  \bibinfo{author}{\bibfnamefont{S.}~\bibnamefont{Boyd}}, in
  \emph{\bibinfo{booktitle}{Recent Advances in Learning and Control}}, edited
  by \bibinfo{editor}{\bibfnamefont{V.}~\bibnamefont{Blondel}},
  \bibinfo{editor}{\bibfnamefont{S.}~\bibnamefont{Boyd}}, \bibnamefont{and}
  \bibinfo{editor}{\bibfnamefont{H.}~\bibnamefont{Kimura}}
  (\bibinfo{publisher}{Springer-Verlag Limited}, \bibinfo{year}{2008}), Lecture
  Notes in Control and Information Sciences, pp. \bibinfo{pages}{95--110},
  \bibinfo{note}{\url{http://stanford.edu/~boyd/graph_dcp.html}}.

\bibitem[{\citenamefont{Werner}(1989)}]{PhysRevA.40.4277}
\bibinfo{author}{\bibfnamefont{R.~F.} \bibnamefont{Werner}},
  \bibinfo{journal}{Phys. Rev. A} \textbf{\bibinfo{volume}{40}},
  \bibinfo{pages}{4277} (\bibinfo{year}{1989}),
  \urlprefix\url{https://link.aps.org/doi/10.1103/PhysRevA.40.4277}.

\bibitem[{\citenamefont{Vollbrecht and
  Werner}(2001)}]{vollbrecht2001entanglement}
\bibinfo{author}{\bibfnamefont{K.~G.~H.} \bibnamefont{Vollbrecht}}
  \bibnamefont{and} \bibinfo{author}{\bibfnamefont{R.~F.}
  \bibnamefont{Werner}}, \bibinfo{journal}{Physical Review A}
  \textbf{\bibinfo{volume}{64}}, \bibinfo{pages}{062307}
  (\bibinfo{year}{2001}).

\bibitem[{\citenamefont{Horodecki and Horodecki}(1999)}]{PhysRevA.59.4206}
\bibinfo{author}{\bibfnamefont{M.}~\bibnamefont{Horodecki}} \bibnamefont{and}
  \bibinfo{author}{\bibfnamefont{P.}~\bibnamefont{Horodecki}},
  \bibinfo{journal}{Phys. Rev. A} \textbf{\bibinfo{volume}{59}},
  \bibinfo{pages}{4206} (\bibinfo{year}{1999}),
  \urlprefix\url{https://link.aps.org/doi/10.1103/PhysRevA.59.4206}.

\bibitem[{\citenamefont{Rains}(1999)}]{PhysRevA.60.179}
\bibinfo{author}{\bibfnamefont{E.~M.} \bibnamefont{Rains}},
  \bibinfo{journal}{Phys. Rev. A} \textbf{\bibinfo{volume}{60}},
  \bibinfo{pages}{179} (\bibinfo{year}{1999}),
  \urlprefix\url{https://link.aps.org/doi/10.1103/PhysRevA.60.179}.

\bibitem[{\citenamefont{Bennett et~al.}(1999)\citenamefont{Bennett, DiVincenzo,
  Mor, Shor, Smolin, and Terhal}}]{bennett1999unextendible}
\bibinfo{author}{\bibfnamefont{C.~H.} \bibnamefont{Bennett}},
  \bibinfo{author}{\bibfnamefont{D.~P.} \bibnamefont{DiVincenzo}},
  \bibinfo{author}{\bibfnamefont{T.}~\bibnamefont{Mor}},
  \bibinfo{author}{\bibfnamefont{P.~W.} \bibnamefont{Shor}},
  \bibinfo{author}{\bibfnamefont{J.~A.} \bibnamefont{Smolin}},
  \bibnamefont{and} \bibinfo{author}{\bibfnamefont{B.~M.}
  \bibnamefont{Terhal}}, \bibinfo{journal}{Physical Review Letters}
  \textbf{\bibinfo{volume}{82}}, \bibinfo{pages}{5385} (\bibinfo{year}{1999}).

\bibitem[{\citenamefont{\ifmmode~\dot{Z}\else \.{Z}\fi{}yczkowski
  et~al.}(1998)\citenamefont{\ifmmode~\dot{Z}\else \.{Z}\fi{}yczkowski,
  Horodecki, Sanpera, and Lewenstein}}]{zyczkowskiVolumeI}
\bibinfo{author}{\bibfnamefont{K.}~\bibnamefont{\ifmmode~\dot{Z}\else
  \.{Z}\fi{}yczkowski}},
  \bibinfo{author}{\bibfnamefont{P.}~\bibnamefont{Horodecki}},
  \bibinfo{author}{\bibfnamefont{A.}~\bibnamefont{Sanpera}}, \bibnamefont{and}
  \bibinfo{author}{\bibfnamefont{M.}~\bibnamefont{Lewenstein}},
  \bibinfo{journal}{Phys. Rev. A} \textbf{\bibinfo{volume}{58}},
  \bibinfo{pages}{883} (\bibinfo{year}{1998}),
  \urlprefix\url{https://link.aps.org/doi/10.1103/PhysRevA.58.883}.

\bibitem[{\citenamefont{\ifmmode~\dot{Z}\else
  \.{Z}\fi{}yczkowski}(1999)}]{zyczkowskiVolumeII}
\bibinfo{author}{\bibfnamefont{K.}~\bibnamefont{\ifmmode~\dot{Z}\else
  \.{Z}\fi{}yczkowski}}, \bibinfo{journal}{Phys. Rev. A}
  \textbf{\bibinfo{volume}{60}}, \bibinfo{pages}{3496} (\bibinfo{year}{1999}),
  \urlprefix\url{https://link.aps.org/doi/10.1103/PhysRevA.60.3496}.

\end{thebibliography}

\end{document}